\documentclass[galaxies,article,accept,pdftex,moreauthors]{Definitions/mdpi} 

\usepackage{xspace}
\usepackage{aas_macros}
\usepackage{soul}
\usepackage{upgreek}

\def\M87{M87$^*$\xspace}
\def\m87{M87$^*$\xspace}
\def\sgra{Sgr A$^*$\xspace}

\firstpage{1} 
\pubvolume{1}
\issuenum{1}
\articlenumber{0}
\pubyear{2022}
\copyrightyear{2022}
\externaleditor{Academic Editor: {Bidzina Kapanadze}}
\datereceived{15 November 2022} 
\dateaccepted{13 December 2022} 
\datepublished{} 
\hreflink{https://doi.org/} %

\Title{Millimeter/Submillimeter VLBI with a Next Generation Large Radio Telescope in the Atacama Desert}

\TitleCitation{Millimeter/Submillimeter VLBI with a Next Generation Large Radio Telescope in the Atacama Desert}

\Author{Kazunori Akiyama $^{1,2,3,}$*\orcidA{}, Jens Kauffmann $^{1}$\orcidB{}, Lynn D. Matthews $^{1}$\orcidC{}, Kotaro Moriyama $^{2,4}$\orcidD{}, \mbox{Shoko Koyama $^{5,6}$\orcidE{}} and Kazuhiro Hada $^{2,7}$\orcidF{}}

\AuthorNames{Kazunori Akiyama, Jens Kauffmann, Lynn D. Matthews, Kotaro Moriyama, Shoko Koyama, Kazuhiro Hada}

\AuthorCitation{Akiyama, K.; Kauffmann, J.; Matthews, L.D.; Moriyama, K.; Koyama, S.; Hada, K.}

\address{%
$^{1}$ \quad Massachusetts Institute of Technology Haystack Observatory, 99 Millstone Rd, Westford, MA 01886, USA; jens.kauffmann@mit.edu (J.K.); lmatthew@haystack.mit.edu (L.D.M.)\\
$^{2}$ \quad Mizusawa VLBI Observatory, National Astronomical Observatory of Japan, 2-12, Hoshigaoka, Mizusawa, Oshu, Iwate 023-0861, Japan; moriyama@itp.uni-frankfurt.de (K.M.); kazuhiro.hada@nao.ac.jp (K.H.)\\
$^{3}$ \quad Black Hole Initiative at Harvard University, 20 Garden Street, Cambridge, MA 02138, USA\\
$^{4}$ \quad Institut f{\" u}r Theoretische Physik, Goethe-Universit{\" a}t Frankfurt, Max-von-Laue-Stra${\beta}$e 1, \mbox{D-60438 Frankfurt am Main,} Germany\\
$^{5}$ \quad Graduate School of Science and Technology, Niigata University, 8050 Ikarashi-nino-cho, Nishi-ku, Niigata 950-2181, Japan; skoyama@create.niigata-u.ac.jp\\ %
$^{6}$ \quad Institute of Astronomy and Astrophysics, Academia Sinica, 11F of Astronomy-Mathematics Building, AS/NTU No. 1, Sec. 4, Roosevelt Rd, Taipei 10617, Taiwan\\
$^{7}$ \quad Department of Astronomical Science, The Graduate University for Advanced Studies (SOKENDAI), \mbox{2-21-1 Osawa,} Mitaka, Tokyo 181-8588, Japan
}

\corres{Correspondence: kakiyama@mit.edu}

\abstract{
The proposed next generation Event Horizon Telescope (ngEHT) concept envisions the imaging of various astronomical sources on scales of microarcseconds in unprecedented detail with at least two orders of magnitude improvement in the image dynamic ranges by extending the Event Horizon Telescope (EHT). A key technical component of ngEHT is the utilization of large aperture telescopes to anchor the entire array, allowing the connection of less sensitive stations through highly sensitive fringe detections to form a dense network across the planet. Here, we introduce two projects for planned next generation large radio telescopes in the 2030s on the Chajnantor Plateau in the Atacama desert in northern Chile, the Large Submillimeter Telescope (LST) and the Atacama Large Aperture Submillimeter Telescope (AtLAST). Both are designed to have a 50-meter diameter and operate at the planned ngEHT frequency bands of 86, 230 and 345\,GHz. A large aperture of 50\,m that is co-located with two existing EHT stations, the Atacama Large Millimeter/Submillimeter Array (ALMA) and the Atacama Pathfinder Experiment (APEX) Telescope in the excellent observing site of the Chajnantor Plateau, will offer excellent capabilities for highly sensitive, multi-frequency, and time-agile millimeter {very long baseline interferometry (VLBI)} observations with accurate data calibration relevant to key science cases of ngEHT. In addition to ngEHT, its unique location in Chile will substantially improve angular resolutions of the planned Next Generation Very Large Array in North America or any future global millimeter VLBI arrays if combined. LST and AtLAST will be a key element enabling transformative science cases with next-generation millimeter/submillimeter VLBI arrays.
}


\begin{document}

\section{Introduction}
With the success of the Event Horizon Telescope\endnote{\url{https://eventhorizontelescope.org/} (accessed on 15 November 2022)} (EHT) \citep{EHTM87PaperI,EHTM87PaperII,EHTM87PaperIII,EHTM87PaperIV,EHTM87PaperV,EHTM87PaperVI,EHTM87PaperVII,EHTM87PaperVIII,EHTSgrAPaperI,EHTSgrAPaperII,EHTSgrAPaperIII,EHTSgrAPaperIV,EHTSgrAPaperV,EHTSgrAPaperVI}, the next generation Event Horizon Telescope\endnote{\url{https://www.ngeht.org/} (accessed on 15 November 2022)} %
(ngEHT) has been proposed as a development concept for the extension of EHT in the 2030s \citep{Doeleman2019a, Doeleman2019b, Raymond2021}. ngEHT aims to extend EHT by adding $\sim$10 new %
stations to {its network of the very long baseline interferometry (VLBI) array} together with overall upgrades in the receiving system, including a significant increase in bandwidth, comparable to the planned Next Generation Very Large Array (ngVLA; \cite{Selina2018, Murphy2018}), along with the capability to perform simultaneous dual- or tri-bands observations \citep{Doeleman2019b, Issaoun2022}. 
Given substantial upgrades in the instruments and VLBI network, ngEHT is anticipated to provide simultaneous multi-frequency imaging at dynamic ranges at least two orders of magnitude better than the current EHT. 
For instance, for \m87, ngEHT is expected to achieve an image dynamic range of >1000 enough to capture a detailed shape of the extended jet emission on scales of thousands Schwarzschild radii, e.g., \citep{Doeleman2019b, Raymond2021}, which was not possible with the EHT 2017 array, which achieved a dynamic range of only $\sim$10 {due to its sparse baseline coverage} \citep{EHTM87PaperIV}.
These unprecedented capabilities allow transformative science cases at extreme high angular resolutions of a few tens of microarcseconds, not only for horizon-scale black hole astrophysics in sources such as \m87 and \sgra, but also in potential other targets for which the array may resolve the horizon-scale emission \citep{Doeleman2019a, Pesce2021} and various compact objects on the sky (see other articles in this special issue).

A key design aspect of the ngEHT array is the use of small and large aperture telescopes to form a dense interferometric network \citep{Doeleman2019a, Doeleman2019b, Raymond2021}. The large-aperture telescopes will work as sensitive anchor stations that will facilitate robust fringe detections across the entire array, whereas the small telescopes will fill up the Fourier coverage of the array and enable high-dynamic-range imaging. The anticipated anchor stations include existing EHT stations such as the Large Millimeter Telescope (LMT), the Atacama Large Millimeter/submillimeter Array (ALMA), and the Northern Extended Millimeter Array (NOEMA), as well as planned additional stations such as the Haystack 37 m Telescope \citep{Kauffmann2022}. 
As the baseline sensitivity is proportional to the geometric mean of the collecting area of the apertures on both ends, the participation of such sensitive facilities is essential for various science cases requiring high-sensitivity observations.

Here, we describe two international projects for planned next generation large radio telescopes in the 2030s, the Large Submillimeter Telescope\endnote{\url{https://en.lstobservatory.org/} (accessed on 15 November 2022)} (LST) \citep{Kawabe2016, Kohno2020} and the Atacama Large Aperture Submillimeter Telescope\endnote{\url{https://www.atlast.uio.no/} (accessed on 15 November 2022)} (AtLAST) \citep{Klaassen2020,Ramasawmy2022}, as a potential anchor stations of next-generation global VLBI arrays. Both projects aim to construct a 50-meter-class radio telescope on the Chajnantor Plateau in the Atacama desert in northern Chile operating at millimeter/submillimeter wavelengths including the planned ngEHT observing frequency bands. The remainder of the paper is constructed as follows. We first describe each project briefly in Section \ref{sec:2}, and then discuss the prospects for having such a large-aperture dish in Atacama for the next generation global submillimeter/millimeter VLBI arrays in Section \ref{sec:3}. Finally, we will make a brief summary and conclusion in Section \ref{sec:4}. %

\section{Planned Large Submillimeter/Millimeter Radio Telescopes in the Atacama Desert \label{sec:2}}
\unskip
\subsection{Large Submillimeter Telescope (LST)}
LST is a planned 50-meter-class single-dish telescope operating at submillimeter and millimeter wavelengths to be constructed on the Chajnantor Plateau in Chile at the same site as ALMA. This project is driven by an international collaboration led by the Japanese radio astronomy community \citep{Kawabe2016, Kohno2020}. The LST concept was originally developed as a next-generation successor of the Nobeyama Radio Observatory (NRO) 45 m telescope \citep{Ukita1994} and the Atacama Submillimeter Telescope Experiment (ASTE) 10 m telescope \citep{Ezawa2008}. LST aims to inherit two major key strengths from these predecessors: a large collection area from the NRO 45 m telescope and the submillimeter capabilities from the ASTE 10-meter telescope. 

\begin{figure}[t] %
 \centering
 \begin{tabular}{cc}
 \includegraphics[height=3.5cm]{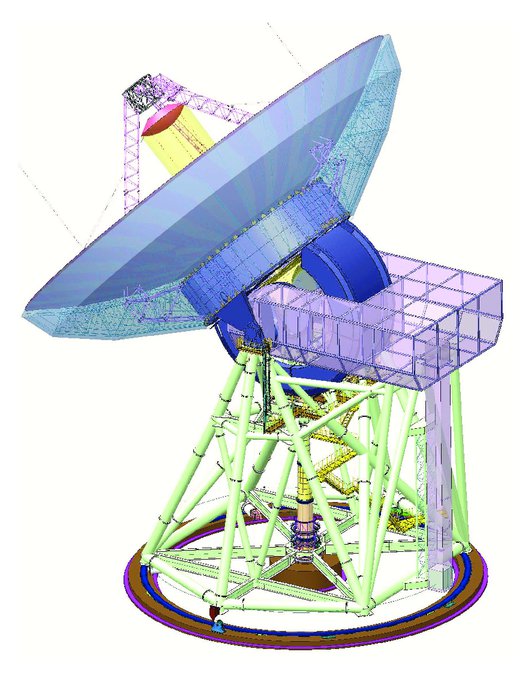}
 \includegraphics[height=3.5cm]{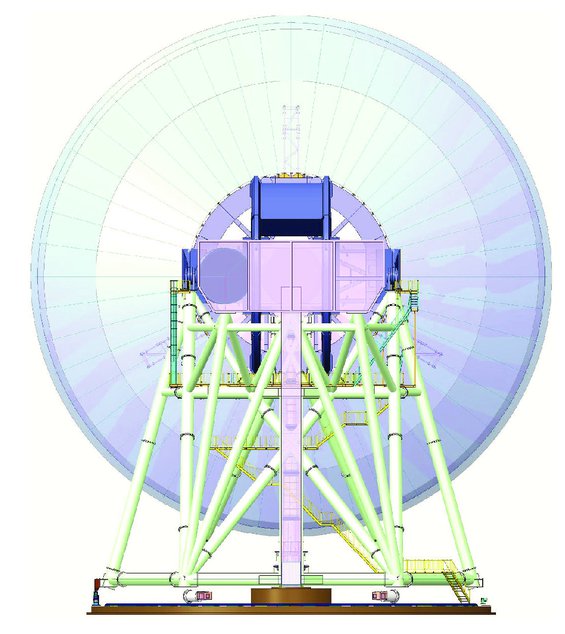} &
 \includegraphics[height=3.5cm]{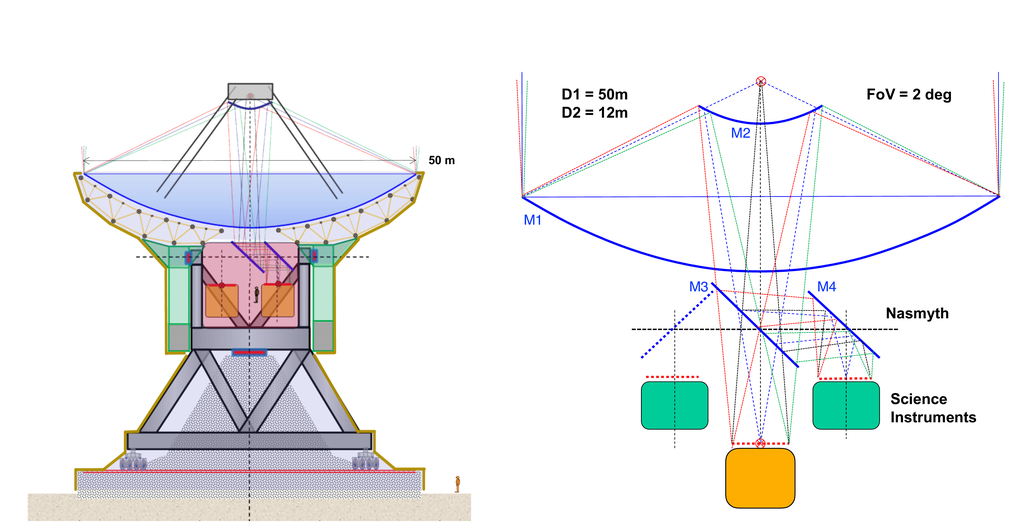}\\
 (\textbf{a}) LST & (\textbf{b}) AtLAST
 \end{tabular}
 \caption{Conceptual designs of (\textbf{a}) LST and (\textbf{b}) AtLAST. %
 Both projects aim a large-aperture telescope with the diameter of $\sim$50 m in Atacama, Chile operating at millimeter and submillimeter wavelengths. The AtLAST concept images are designed before the ongoing European Union funded Horizon 2020 research and innovation programme. The current optics design has evolved significantly under the programme (see Section \ref{sec:2.2}).
 The LST concept images are in courtesy of Mitsubishi Electric Corporation (MELCO), and adapted with permission from Ref. \cite{Kawabe2016}; Copyright 2016, Society of Photo‑Optical Instrumentation Engineers (SPIE).
 The AtLAST images are adapted with permission from Ref. \cite{Klaassen2020}; Copyright 2020, SPIE.
 }
 \label{fig:designs}
\end{figure}

{The current key conceptual design and major specifications are described in {Kawabe et al. \cite{Kawabe2016}.} LST is planned to have a 50-meter-diameter dish (see Figure \ref{fig:designs}a) with a high surface precision (45 $\upmu$m rms) designed for wide-area imaging and spectroscopic surveys with a field-of-view of $\sim$1$^\circ$ primarily focusing on the 70–420\,GHz frequency range. }

In the current LST design, the targeted wide field of view is enabled by adopting a Ritchey--Chr{\' e}tien (RC) system for optics \citep{Kawabe2016}.
The project further aims to have a capability for frequencies up to 1 THz using an inner high-precision surface.
To establish a high-precision surface with a large collecting area, LST plans to implement a millimetric adaptive optics system based on real-time sensing of the surface with a dedicated millimeter wavefront sensor, e.g., \citep{Tamura2020}.
Key science cases enabled by the highly-sensitive large-aperture of LST, briefly summarized {in \citet{Kawabe2016}}, include black hole astrophysics with high-sensitive millimeter/submillimeter VLBI involving the LST as an anchor station for global arrays.

The recent progress and the near-future prospects of the project are summarized {{in} \citet{Kohno2020}. }
The LST project is formally listed as a large academic project in the astronomy and astrophysics division in the Master Plan 2020 led by the Science Council of Japan, the latest series of surveys designed to review and maintain the list of high-priority large academic research projects in Japan every three or four years. 
The LST project anticipates the merger of the project into the AtLAST project in the mid-2020s. Collaborative studies would be undertaken under the ongoing AtLAST design study program after resolving inconsistencies in telescope specifications between the two projects. 

\subsection{Atacama Large Aperture Submillimeter Telescope (AtLAST)\label{sec:2.2}}
Just like LST, AtLAST is planned to be built on the Chajnantor Plateau, in close proximity to ALMA. An ongoing design study for the observatory, supported by the European Union’s Horizon 2020 research and innovation program, seeks to further refine the details in the 2021--2024 time frame. This section primarily summarizes the specifications as known at the beginning of the design study \citep{Klaassen2020} and the current key science drivers \citep{Ramasawmy2022}.

The telescope will be sited at an altitude between 5100 and 5500 m, depending on the specific location chosen for construction. The project seeks to deliver a dish with a diameter of 50 m and a very large field of view ($\approx$2$^{\circ}$) achieved by an optics system involving a large secondary mirror (see Figure \ref{fig:designs}b for its early concept design). The latter will enable a new generation of astrophysical experiments that cannot be pursued otherwise. The goal is to provide access to frequencies $\gtrsim{}850 \rm{}GHz$, resulting in a desired dish surface accuracy of 20--$25 \upmu{}\rm{}m$. An active surface would be employed to achieve this precision. We note that the AtLAST design shown in Figure \ref{fig:designs}b, adapted {from \citet{Klaassen2020},} has matured through the European Union funded Horizon 2020 research and innovation program into a 3-mirror, hybrid Nasmyth-like design that will be able to host two 2 degree wide field-of-view instruments located along the elevation axis, as well as several additional smaller instruments with up to 1 degree diameter fields-of-view, located off axis, without requiring additional external re-imaging optics.

AtLAST would enable breakthroughs in several domains of astrophysics. 
These include studies of molecular clouds in the Milky Way, galaxies and their formation over cosmic time, as well as the evolution of galaxy clusters. These objects can be investigated using line emission from molecules and atoms, continuum emission from dust, and by employing the broadband spectral signature of the Sunyaev--Zeldovich (SZ) effect that probes the ionized medium in and between galaxies. AtLAST will be optimized for wide-field surveys, but it is conceived to be a multi-role observatory that would be open to PI-driven research projects.

\section{Prospects for Millimeter/Submillimeter VLBI with LST and AtLAST \label{sec:3}}
{A next-generation large millimeter/submillimeter telescope in the Atacama desert, if realized, could play a vital role in global millimeter/submillimeter VLBI observations in multiple ways. 
An obvious strength is sensitivity, for instance, a 50 m diameter telescope with an aperture efficiency of $\sim$50\,\% and a system noise temperature of $\sim$100\,K at 230\,GHz anticipated for the site, e.g., \citep{EHTM87PaperII} will achieve the system-equivalent flux density (SEFD) of $\sim$300\,Jy. 
It is better than the anticipated sensitivity of the phased array of the Northern Extended Millimeter Array (NOEMA) with the SEFD of $\sim$700\,Jy, e.g., \citep{EHTM87PaperII}, and orders of magnitude better than the typical sensitivities of the existing and anticipated EHT or ngEHT stations with SEFDs of $\sim$1000--20,000\,Jy \citep{EHTM87PaperII,EHTM87PaperIII,EHTSgrAPaperII}. 
The telescope is expected to have a competitive sensitivity of $\sim$30--40\,\% of ALMA with the SEFD of $\sim$100\,Jy\endnote{The current ALMA 2030 roadmap aims to improve the sensitivity by the overall upgrade in the frontend and also the correlator, which would provide a better SEFD.} \citep{EHTM87PaperII, EHTM87PaperIII, EHTSgrAPaperII} which is the highest sensitivity among the submillimeter VLBI stations anticipated in 2030s. }

{The location of the planned site in the Atacama provides additional benefits to global VLBI observations. First, the telescopes in the Atacama provide very long, intercontinental baselines, especially in the north-south directions to North American, European, and Pacific stations. The Atacama baselines have been providing substantial improvements in the angular resolution of the Global Millimeter VLBI Array (GMVA), e.g., \citep{Issaoun2019, Issaoun2021, Okino2021, Zhao2022} and were essential to resolving the shadows of the supermassive black holes \m87 and \sgra with the EHT, e.g., \citep{EHTM87PaperIV}. 
The 50-meter telescope in the Atacama desert will be a key anchor station to secure the detection of fringes on intercontinental baselines and enhance the sensitivity of the overall array.
Second, the telescope shares its site with other submillimeter/millimeter facilities such as ALMA and APEX, providing redundant baselines as well as the intra-site baselines to the entire ALMA array. 
Inteferometric measurements on baselines from redundant stations allow accurate and precise absolute calibrations of interferometric data, e.g., \citep{Fish2011, Johnson2015, EHTM87PaperIII, EHTM87PaperVII} which are critical for both total-intensity and polarimetric imaging.}

With its competitive sensitivity, a large-aperture single-dish telescope has potential strengths for global VLBI observations over phased stations such as the colocated ALMA. These strengths benefit from simpler instrumentation and observing logistics required for a single dish telescope to be a VLBI station. 

\subsection{{Observations of Fainter Sources}}
With its competitive sensitivity, a large aperture single-dish telescope has a unique strength that can broaden the number of faint target sources that may not be observable sorely with phased array stations. 
Phased arrays often need the target sources to be bright enough for active phasing, or alternatively to have bright and compact phase calibrators nearby for passive phasing, for instance, the current ALMA phasing system has a limit on the total flux density of $\sim$500\,mJy for the target sources or phase calibrators with a separation only within several degrees\endnote{see e.g., ALMA Cycle 8 Proposer's Guide: \url{https://almascience.nrao.edu/documents-and-tools/cycle8/alma-proposers-guide} (accessed on 15 November 2022)}. 
Its competitive sensitivity without the need for phasing will be critical for faint science targets that may not necessarily have appropriate phase calibrators.

\subsection{{Simultaneous Multi-Frequency Observations}}
{Although the anticipated sensitivity is a few times lower than that of ALMA, a 50 m single dish in Atacama may have unique strengths compared to ALMA in frequency agility while maintaining its high sensitivity. 
An important specification of ngEHT is a capability of simultaneous multi-frequency observations \citep{Doeleman2019b}, allowing the order-of-magnitude improvements in sensitivity with frequency phase transfer (FPT) \mbox{techniques \citep{Asaki1996, Dodson2009, Rioja2011, Rioja2022}}, as well as enabling various science cases based on simultaneous measurements of full-polarization spectra and Faraday-rotations of various sources. 
Although ngEHT originally aimed to have simultaneous dual-band receiving at 230 and 340 \ GHz \citep{Doeleman2019b}, the project is now pursuing enabling simultaneous tri-band observations at 86, 230 and 345 GHz \citep{Issaoun2022}. }

While a capability of simultaneous multi-frequency observations can be realized by the implementation of a dedicated receiving system for single-dish telescopes, it is more logistically and operationally complicated for phased ALMA. 
To keep its sensitivity at each frequency, it will need installations of such receiving systems across the array. 
Alternatively, simultaneous multi-band observations may be enabled by splitting the entire array into subarrays operating at single frequencies, which will compensate the sensitivity at each frequency due to the reduced synthesized aperture.
The current 2030 roadmap of ALMA is targeting the latter approach by implementing subarraying capabilities instead of simultaneous dual- or tri-frequency operations with a single array \citep{Carlson2020}, indicating that the next-generation large telescope in the Atacama has a strong potential to have a comparable or even better sensitivity than ALMA if a dedicated receiving system is implemented for simultaneous multi-band observations.

The benefits of having a large single dish for simultaneous multi-band observations are not limited to high sensitivity. A strong advantage of simultaneous multi-frequency observations with a single dish over a subarray is that the signals at multiple frequencies natively share the same atmospheric line of sight. This is important for the application of FPT techniques that allow much longer phase coherence and integration of interferometric fringes at higher frequencies by solving the atmospheric phase delays at lower frequencies. Although the application of the FPT techniques to subarrays is possible by computationally aligning the phase center of subarrays to the same location, it will complicate the signal processing and may cause the additional loss in the phase coherence and/or systematic errors in FPT-applied data.

\subsection{{Time-Domain Science}}
Another potential strength of a single-dish sensitive telescope in the Atacama is time agility, which will allow various time-domain science requiring monitoring observations over weeks and months proposed for ngEHT and also other global arrays such as ngVLA and a next-generation GMVA.
Phased arrays are known to be more impacted by windy, turbulent weather, and the resulting worse atmospheric phase coherence, since the atmospheric delay in the radio signal received in each antenna needs to be corrected accurately in real time, e.g., \citep{Matthews2018}.
For example, the phasing efficiency of ALMA is reported to be consistently poor at 345\,GHz when wind speeds exceed $\sim$10\,m/s regardless of the amount of the precipitable water vapor (PWV) \citep{Crew2022}. 
While a single-dish telescope may be affected by windy weather through, for instance, losses in focus and pointing efficiency, it may provide more robust observing capabilities during periods of atmospheric stability, and ultimately enable expansion of suitable observing windows.

\section{Summary \label{sec:4}}
In this article, we have introduced two planned projects for a next-generation submillimeter/millimeter single-dish telescope, LST and AtLAST, both aiming a 50 m-class telescope in the Atacama Desert of Chile, sharing the site with two existing submillimeter/millimeter facilities, APEX and ALMA. The two projects are currently anticipated to be merged in the next several years as a result of ongoing study programs.
We further discussed the strong and unique benefits of having such a large aperture dish in the Atacama as part of the planned next-generation of millimeter/sub-millimeter VLBI arrays including ngEHT and ngVLA. 
A large-aperture single-dish telescope at this location, with its excellent observing conditions, will play a vital role as a high-sensitivity key anchor station capable of significantly improving the sensitivity and angular resolution of planned global VLBI arrays will potentially have better time and frequency agility than phased ALMA. 
LST, AtLAST, or a future merged telescope thus have strong potential to be a key element for next generation millimeter/submillimeter VLBI arrays and allow them to achieve a range of new science goals requiring highly sensitive, multi-frequency, time-agile observations.

\vspace{6pt}

\authorcontributions{Writing---original draft preparation, {K.A. and J.K.}; writing---review and editing, all authors (K.A., J.K., L.D.M, K.M., S.K., and K.H.). All authors contributed equally to this work. All authors have read and agreed to the published version of the manuscript.} 

\funding{This work is financially supported by grants from the National Science Foundation (NSF; AST-1440254, AST-1614868, AST-1935980, AST-2034306). 
K.A. has been also financially supported by other NSF grants (OMA-2029670, AST-2107681, AST-2132700). 
AtLAST has received funding from the European Union’s Horizon 2020 research and innovation programme under grant agreement No. 951815.
}

\institutionalreview{Not applicable}

\informedconsent{Not applicable}

\dataavailability{Not applicable}

\acknowledgments{We thank Pamela Klaassen, Tony Mroczkowski, Claudia Cicone, and Colin J. Lonsdale for helpful feedback to improve the article. We appreciate the internal reviewer of this article from the EHT collaboration, Lindy Blackburn for helpful and constructive suggestions on the article.}

\conflictsofinterest{The authors declare no conflict of interest.}

\abbreviations{Abbreviations}{
The following abbreviations are used in this manuscript:\\

\noindent 
\begin{tabular}{@{}ll}
ALMA & Atacama Large Millimeter/submillimeter Array\\
AtLAST & Atacama Large Aperture Submillimeter Telescope\\
APEX & Atacama Pathfinder Experiment\\
EHT & Event Horizon Telescope\\
GMVA & Global Millimeter VLBI Array\\
LMT & Large Millimeter Telescope\\
LST & Large Submillimeter Telescope\\
ngEHT & next generation Event Horizon Telescope\\
ngVLA & next generation Very Large Array\\
NOEMA & Nothern Extended Millimeter Array\\
VLBI & Very Long Baseline Interferometry
\end{tabular}
}

\begin{adjustwidth}{-\extralength}{0cm}
\printendnotes[custom]
\reftitle{References}

\end{adjustwidth}

\begin{thebibliography}{999}

\bibitem[{Event Horizon Telescope Collaboration}
 {et al.}(2019{\natexlab{a}}){Event Horizon Telescope Collaboration},
 {Akiyama}, {Alberdi}, {Alef}, {Asada}, {Azulay}, {Baczko}, {Ball},
 {Balokovi{\'c}}, {Barrett}, {Bintley}, {Blackburn}, {Boland}, {Bouman},
 {Bower}, {Bremer}, {Brinkerink}, {Brissenden}, {Britzen}, {Broderick},
 {Broguiere}, {Bronzwaer}, {Byun}, {Carlstrom}, {Chael}, {Chan}, {Chatterjee},
 {Chatterjee}, {Chen}, {Chen}, {Cho}, {Christian}, {Conway}, {Cordes}, {Crew},
 {Cui}, {Davelaar}, {De Laurentis}, {Deane}, {Dempsey}, {Desvignes}, {Dexter},
 {Doeleman}, {Eatough}, {Falcke}, {Fish}, {Fomalont}, {Fraga-Encinas},
 {Freeman}, {Friberg}, {Fromm}, {G{\'o}mez}, {Galison}, {Gammie},
 {Garc{\'\i}a}, {Gentaz}, {Georgiev}, {Goddi}, {Gold}, {Gu}, {Gurwell},
 {Hada}, {Hecht}, {Hesper}, {Ho}, {Ho}, {Honma}, {Huang}, {Huang}, {Hughes},
 {Ikeda}, {Inoue}, {Issaoun}, {James}, {Jannuzi}, {Janssen}, {Jeter}, {Jiang},
 {Johnson}, {Jorstad}, {Jung}, {Karami}, {Karuppusamy}, {Kawashima},
 {Keating}, {Kettenis}, {Kim}, {Kim}, {Kim}, {Kino}, {Koay}, {Koch}, {Koyama},
 {Kramer}, {Kramer}, {Krichbaum}, {Kuo}, {Lauer}, {Lee}, {Li}, {Li},
 {Lindqvist}, {Liu}, {Liuzzo}, {Lo}, {Lobanov}, {Loinard}, {Lonsdale}, {Lu},
 {MacDonald}, {Mao}, {Markoff}, {Marrone}, {Marscher}, {Mart{\'\i}-Vidal},
 {Matsushita}, {Matthews}, {Medeiros}, {Menten}, {Mizuno}, {Mizuno}, {Moran},
 {Moriyama}, {Moscibrodzka}, {M{\"u}ller}, {Nagai}, {Nagar}, {Nakamura},
 {Narayan}, {Narayanan}, {Natarajan}, {Neri}, {Ni}, {Noutsos}, {Okino},
 {Olivares}, {Ortiz-Le{\'o}n}, {Oyama}, {{\"O}zel}, {Palumbo}, {Patel}, {Pen},
 {Pesce}, {Pi{\'e}tu}, {Plambeck}, {PopStefanija}, {Porth}, {Prather},
 {Preciado-L{\'o}pez}, {Psaltis}, {Pu}, {Ramakrishnan}, {Rao}, {Rawlings},
 {Raymond}, {Rezzolla}, {Ripperda}, {Roelofs}, {Rogers}, {Ros}, {Rose},
 {Roshanineshat}, {Rottmann}, {Roy}, {Ruszczyk}, {Ryan}, {Rygl},
 {S{\'a}nchez}, {S{\'a}nchez-Arguelles}, {Sasada}, {Savolainen}, {Schloerb},
 {Schuster}, {Shao}, {Shen}, {Small}, {Sohn}, {SooHoo}, {Tazaki}, {Tiede},
 {Tilanus}, {Titus}, {Toma}, {Torne}, {Trent}, {Trippe}, {Tsuda}, {van
 Bemmel}, {van Langevelde}, {van Rossum}, {Wagner}, {Wardle}, {Weintroub},
 {Wex}, {Wharton}, {Wielgus}, {Wong}, {Wu}, {Young}, {Young}, {Younsi},
 {Yuan}, {Yuan}, {Zensus}, {Zhao}, {Zhao}, {Zhu}, {Algaba}, {Allardi},
 {Amestica}, {Anczarski}, {Bach}, {Baganoff}, {Beaudoin}, {Benson},
 {Berthold}, {Blanchard}, {Blundell}, {Bustamente}, {Cappallo},
 {Castillo-Dom{\'\i}nguez}, {Chang}, {Chang}, {Chang}, {Chen}, {Chilson},
 {Chuter}, {C{\'o}rdova Rosado}, {Coulson}, {Crawford}, {Crowley}, {David},
 {Derome}, {Dexter}, {Dornbusch}, {Dudevoir}, {Dzib}, {Eckart}, {Eckert},
 {Erickson}, {Everett}, {Faber}, {Farah}, {Fath}, {Folkers}, {Forbes},
 {Freund}, {G{\'o}mez-Ruiz}, {Gale}, {Gao}, {Geertsema}, {Graham}, {Greer},
 {Grosslein}, {Gueth}, {Haggard}, {Halverson}, {Han}, {Han}, {Hao},
 {Hasegawa}, {Henning}, {Hern{\'a}ndez-G{\'o}mez}, {Herrero-Illana},
 {Heyminck}, {Hirota}, {Hoge}, {Huang}, {Impellizzeri}, {Jiang}, {Kamble},
 {Keisler}, {Kimura}, {Kono}, {Kubo}, {Kuroda}, {Lacasse}, {Laing}, {Leitch},
 {Li}, {Lin}, {Liu}, {Liu}, {Lu}, {Marson}, {Martin-Cocher}, {Massingill},
 {Matulonis}, {McColl}, {McWhirter}, {Messias}, {Meyer-Zhao}, {Michalik},
 {Monta{\ n}a}, {Montgomerie}, {Mora-Klein}, {Muders}, {Nadolski}, {Navarro},
 {Neilsen}, {Nguyen}, {Nishioka}, {Norton}, {Nowak}, {Nystrom}, {Ogawa},
 {Oshiro}, {Oyama}, {Parsons}, {Paine}, {Pe{\ n}alver}, {Phillips}, {Poirier},
 {Pradel}, {Primiani}, {Raffin}, {Rahlin}, {Reiland}, {Risacher}, {Ruiz},
 {S{\'a}ez-Mada{\'\i}n}, {Sassella}, {Schellart}, {Shaw}, {Silva}, {Shiokawa},
 {Smith}, {Snow}, {Souccar}, {Sousa}, {Sridharan}, {Srinivasan}, {Stahm},
 {Stark}, {Story}, {Timmer}, {Vertatschitsch}, {Walther}, {Wei}, {Whitehorn},
 {Whitney}, {Woody}, {Wouterloot}, {Wright}, {Yamaguchi}, {Yu}, {Zeballos},
 {Zhang}, and {Ziurys}]{EHTM87PaperI}
{Event Horizon Telescope Collaboration}; {Akiyama}, K.; {Alberdi}, A.; {Alef},
 W.; {Asada}, K.; {Azulay}, R.; {Baczko}, A.K.; {Ball}, D.; {Balokovi{\'c}},
 M.; {Barrett}, J.; et al.
\newblock {First M87 Event Horizon Telescope Results. I. The Shadow of the
 Supermassive Black Hole}.
\newblock {\em Astrophys. J. Lett.} {\bf 2019}, {\em 875}, L1. 
\newblock {{https://doi.org/10.3847/2041-8213/ab0ec7}}.

\bibitem[{Event Horizon Telescope Collaboration}
 {et al.}(2019{\natexlab{b}}){Event Horizon Telescope Collaboration},
 {Akiyama}, {Alberdi}, {Alef}, {Asada}, {Azulay}, {Baczko}, {Ball},
 {Balokovi{\'c}}, {Barrett}, {Bintley}, {Blackburn}, {Boland}, {Bouman},
 {Bower}, {Bremer}, {Brinkerink}, {Brissenden}, {Britzen}, {Broderick},
 {Broguiere}, {Bronzwaer}, {Byun}, {Carlstrom}, {Chael}, {Chan}, {Chatterjee},
 {Chatterjee}, {Chen}, {Chen}, {Cho}, {Christian}, {Conway}, {Cordes}, {Crew},
 {Cui}, {Davelaar}, {De Laurentis}, {Deane}, {Dempsey}, {Desvignes}, {Dexter},
 {Doeleman}, {Eatough}, {Falcke}, {Fish}, {Fomalont}, {Fraga-Encinas},
 {Friberg}, {Fromm}, {G{\'o}mez}, {Galison}, {Gammie}, {Garc{\'\i}a},
 {Gentaz}, {Georgiev}, {Goddi}, {Gold}, {Gu}, {Gurwell}, {Hada}, {Hecht},
 {Hesper}, {Ho}, {Ho}, {Honma}, {Huang}, {Huang}, {Hughes}, {Ikeda}, {Inoue},
 {Issaoun}, {James}, {Jannuzi}, {Janssen}, {Jeter}, {Jiang}, {Johnson},
 {Jorstad}, {Jung}, {Karami}, {Karuppusamy}, {Kawashima}, {Keating},
 {Kettenis}, {Kim}, {Kim}, {Kim}, {Kino}, {Koay}, {Koch}, {Koyama}, {Kramer},
 {Kramer}, {Krichbaum}, {Kuo}, {Lauer}, {Lee}, {Li}, {Li}, {Lindqvist}, {Liu},
 {Liuzzo}, {Lo}, {Lobanov}, {Loinard}, {Lonsdale}, {Lu}, {MacDonald}, {Mao},
 {Markoff}, {Marrone}, {Marscher}, {Mart{\'\i}-Vidal}, {Matsushita},
 {Matthews}, {Medeiros}, {Menten}, {Mizuno}, {Mizuno}, {Moran}, {Moriyama},
 {Moscibrodzka}, {M{\"u}ller}, {Nagai}, {Nagar}, {Nakamura}, {Narayan},
 {Narayanan}, {Natarajan}, {Neri}, {Ni}, {Noutsos}, {Okino}, {Olivares},
 {Ortiz-Le{\'o}n}, {Oyama}, {{\"O}zel}, {Palumbo}, {Patel}, {Pen}, {Pesce},
 {Pi{\'e}tu}, {Plambeck}, {PopStefanija}, {Porth}, {Prather},
 {Preciado-L{\'o}pez}, {Psaltis}, {Pu}, {Ramakrishnan}, {Rao}, {Rawlings},
 {Raymond}, {Rezzolla}, {Ripperda}, {Roelofs}, {Rogers}, {Ros}, {Rose},
 {Roshanineshat}, {Rottmann}, {Roy}, {Ruszczyk}, {Ryan}, {Rygl},
 {S{\'a}nchez}, {S{\'a}nchez-Arguelles}, {Sasada}, {Savolainen}, {Schloerb},
 {Schuster}, {Shao}, {Shen}, {Small}, {Sohn}, {SooHoo}, {Tazaki}, {Tiede},
 {Tilanus}, {Titus}, {Toma}, {Torne}, {Trent}, {Trippe}, {Tsuda}, {van
 Bemmel}, {van Langevelde}, {van Rossum}, {Wagner}, {Wardle}, {Weintroub},
 {Wex}, {Wharton}, {Wielgus}, {Wong}, {Wu}, {Young}, {Young}, {Younsi},
 {Yuan}, {Yuan}, {Zensus}, {Zhao}, {Zhao}, {Zhu}, {Algaba}, {Allardi},
 {Amestica}, {Bach}, {Beaudoin}, {Benson}, {Berthold}, {Blanchard},
 {Blundell}, {Bustamente}, {Cappallo}, {Castillo-Dom{\'\i}nguez}, {Chang},
 {Chang}, {Chang}, {Chen}, {Chilson}, {Chuter}, {C{\'o}rdova Rosado},
 {Coulson}, {Crawford}, {Crowley}, {David}, {Derome}, {Dexter}, {Dornbusch},
 {Dudevoir}, {Dzib}, {Eckert}, {Erickson}, {Everett}, {Faber}, {Farah},
 {Fath}, {Folkers}, {Forbes}, {Freund}, {G{\'o}mez-Ruiz}, {Gale}, {Gao},
 {Geertsema}, {Graham}, {Greer}, {Grosslein}, {Gueth}, {Halverson}, {Han},
 {Han}, {Hao}, {Hasegawa}, {Henning}, {Hern{\'a}ndez-G{\'o}mez},
 {Herrero-Illana}, {Heyminck}, {Hirota}, {Hoge}, {Huang}, {Impellizzeri},
 {Jiang}, {Kamble}, {Keisler}, {Kimura}, {Kono}, {Kubo}, {Kuroda}, {Lacasse},
 {Laing}, {Leitch}, {Li}, {Lin}, {Liu}, {Liu}, {Lu}, {Marson},
 {Martin-Cocher}, {Massingill}, {Matulonis}, {McColl}, {McWhirter}, {Messias},
 {Meyer-Zhao}, {Michalik}, {Monta{\ n}a}, {Montgomerie}, {Mora-Klein},
 {Muders}, {Nadolski}, {Navarro}, {Nguyen}, {Nishioka}, {Norton}, {Nystrom},
 {Ogawa}, {Oshiro}, {Oyama}, {Padin}, {Parsons}, {Paine}, {Pe{\ n}alver},
 {Phillips}, {Poirier}, {Pradel}, {Primiani}, {Raffin}, {Rahlin}, {Reiland},
 {Risacher}, {Ruiz}, {S{\'a}ez-Mada{\'\i}n}, {Sassella}, {Schellart}, {Shaw},
 {Silva}, {Shiokawa}, {Smith}, {Snow}, {Souccar}, {Sousa}, {Sridharan},
 {Srinivasan}, {Stahm}, {Stark}, {Story}, {Timmer}, {Vertatschitsch},
 {Walther}, {Wei}, {Whitehorn}, {Whitney}, {Woody}, {Wouterloot}, {Wright},
 {Yamaguchi}, {Yu}, {Zeballos}, and {Ziurys}]{EHTM87PaperII}
{Event Horizon Telescope Collaboration}; {Akiyama}, K.; {Alberdi}, A.; {Alef},
 W.; {Asada}, K.; {Azulay}, R.; {Baczko}, A.K.; {Ball}, D.; {Balokovi{\'c}},
 M.; {Barrett}, J.; et al.
\newblock {First M87 Event Horizon Telescope Results. II. Array and
 Instrumentation}.
\newblock {\em Astrophys. J. Lett.} {\bf 2019}, {\em 875}, L2. 
\newblock {{https://doi.org/10.3847/2041-8213/ab0c96}}.

\bibitem[{Event Horizon Telescope Collaboration}
 {et al.}(2019{\natexlab{c}}){Event Horizon Telescope Collaboration},
 {Akiyama}, {Alberdi}, {Alef}, {Asada}, {Azulay}, {Baczko}, {Ball},
 {Balokovi{\'c}}, {Barrett}, {Bintley}, {Blackburn}, {Boland}, {Bouman},
 {Bower}, {Bremer}, {Brinkerink}, {Brissenden}, {Britzen}, {Broderick},
 {Broguiere}, {Bronzwaer}, {Byun}, {Carlstrom}, {Chael}, {Chan}, {Chatterjee},
 {Chatterjee}, {Chen}, {Chen}, {Cho}, {Christian}, {Conway}, {Cordes}, {Crew},
 {Cui}, {Davelaar}, {De Laurentis}, {Deane}, {Dempsey}, {Desvignes}, {Dexter},
 {Doeleman}, {Eatough}, {Falcke}, {Fish}, {Fomalont}, {Fraga-Encinas},
 {Friberg}, {Fromm}, {G{\'o}mez}, {Galison}, {Gammie}, {Garc{\'\i}a},
 {Gentaz}, {Georgiev}, {Goddi}, {Gold}, {Gu}, {Gurwell}, {Hada}, {Hecht},
 {Hesper}, {Ho}, {Ho}, {Honma}, {Huang}, {Huang}, {Hughes}, {Ikeda}, {Inoue},
 {Issaoun}, {James}, {Jannuzi}, {Janssen}, {Jeter}, {Jiang}, {Johnson},
 {Jorstad}, {Jung}, {Karami}, {Karuppusamy}, {Kawashima}, {Keating},
 {Kettenis}, {Kim}, {Kim}, {Kim}, {Kino}, {Koay}, {Koch}, {Koyama}, {Kramer},
 {Kramer}, {Krichbaum}, {Kuo}, {Lauer}, {Lee}, {Li}, {Li}, {Lindqvist}, {Liu},
 {Liuzzo}, {Lo}, {Lobanov}, {Loinard}, {Lonsdale}, {Lu}, {MacDonald}, {Mao},
 {Markoff}, {Marrone}, {Marscher}, {Mart{\'\i}-Vidal}, {Matsushita},
 {Matthews}, {Medeiros}, {Menten}, {Mizuno}, {Mizuno}, {Moran}, {Moriyama},
 {Moscibrodzka}, {M{\"u}ller}, {Nagai}, {Nagar}, {Nakamura}, {Narayan},
 {Narayanan}, {Natarajan}, {Neri}, {Ni}, {Noutsos}, {Okino}, {Olivares},
 {Ortiz-Le{\'o}n}, {Oyama}, {{\"O}zel}, {Palumbo}, {Patel}, {Pen}, {Pesce},
 {Pi{\'e}tu}, {Plambeck}, {PopStefanija}, {Porth}, {Prather},
 {Preciado-L{\'o}pez}, {Psaltis}, {Pu}, {Ramakrishnan}, {Rao}, {Rawlings},
 {Raymond}, {Rezzolla}, {Ripperda}, {Roelofs}, {Rogers}, {Ros}, {Rose},
 {Roshanineshat}, {Rottmann}, {Roy}, {Ruszczyk}, {Ryan}, {Rygl},
 {S{\'a}nchez}, {S{\'a}nchez-Arguelles}, {Sasada}, {Savolainen}, {Schloerb},
 {Schuster}, {Shao}, {Shen}, {Small}, {Sohn}, {SooHoo}, {Tazaki}, {Tiede},
 {Tilanus}, {Titus}, {Toma}, {Torne}, {Trent}, {Trippe}, {Tsuda}, {van
 Bemmel}, {van Langevelde}, {van Rossum}, {Wagner}, {Wardle}, {Weintroub},
 {Wex}, {Wharton}, {Wielgus}, {Wong}, {Wu}, {Young}, {Young}, {Younsi},
 {Yuan}, {Yuan}, {Zensus}, {Zhao}, {Zhao}, {Zhu}, {Cappallo}, {Farah},
 {Folkers}, {Meyer-Zhao}, {Michalik}, {Nadolski}, {Nishioka}, {Pradel},
 {Primiani}, {Souccar}, {Vertatschitsch}, and {Yamaguchi}]{EHTM87PaperIII}
{Event Horizon Telescope Collaboration}.; {Akiyama}, K.; {Alberdi}, A.; {Alef},
 W.; {Asada}, K.; {Azulay}, R.; {Baczko}, A.K.; {Ball}, D.; {Balokovi{\'c}},
 M.; {Barrett}, J.; et al.
\newblock {First M87 Event Horizon Telescope Results. III. Data Processing and
 Calibration}.
\newblock {\em Astrophys. J. Lett.} {\bf 2019}, {\em 875}, L3. 
\newblock {{https://doi.org/10.3847/2041-8213/ab0c57}}.

\bibitem[{Event Horizon Telescope Collaboration}
 {et al.}(2019{\natexlab{d}}){Event Horizon Telescope Collaboration},
 {Akiyama}, {Alberdi}, {Alef}, {Asada}, {Azulay}, {Baczko}, {Ball},
 {Balokovi{\'c}}, {Barrett}, {Bintley}, {Blackburn}, {Boland}, {Bouman},
 {Bower}, {Bremer}, {Brinkerink}, {Brissenden}, {Britzen}, {Broderick},
 {Broguiere}, {Bronzwaer}, {Byun}, {Carlstrom}, {Chael}, {Chan}, {Chatterjee},
 {Chatterjee}, {Chen}, {Chen}, {Cho}, {Christian}, {Conway}, {Cordes}, {Crew},
 {Cui}, {Davelaar}, {De Laurentis}, {Deane}, {Dempsey}, {Desvignes}, {Dexter},
 {Doeleman}, {Eatough}, {Falcke}, {Fish}, {Fomalont}, {Fraga-Encinas},
 {Freeman}, {Friberg}, {Fromm}, {G{\'o}mez}, {Galison}, {Gammie},
 {Garc{\'\i}a}, {Gentaz}, {Georgiev}, {Goddi}, {Gold}, {Gu}, {Gurwell},
 {Hada}, {Hecht}, {Hesper}, {Ho}, {Ho}, {Honma}, {Huang}, {Huang}, {Hughes},
 {Ikeda}, {Inoue}, {Issaoun}, {James}, {Jannuzi}, {Janssen}, {Jeter}, {Jiang},
 {Johnson}, {Jorstad}, {Jung}, {Karami}, {Karuppusamy}, {Kawashima},
 {Keating}, {Kettenis}, {Kim}, {Kim}, {Kim}, {Kino}, {Koay}, {Koch}, {Koyama},
 {Kramer}, {Kramer}, {Krichbaum}, {Kuo}, {Lauer}, {Lee}, {Li}, {Li},
 {Lindqvist}, {Liu}, {Liuzzo}, {Lo}, {Lobanov}, {Loinard}, {Lonsdale}, {Lu},
 {MacDonald}, {Mao}, {Markoff}, {Marrone}, {Marscher}, {Mart{\'\i}-Vidal},
 {Matsushita}, {Matthews}, {Medeiros}, {Menten}, {Mizuno}, {Mizuno}, {Moran},
 {Moriyama}, {Moscibrodzka}, {M{\"u}ller}, {Nagai}, {Nagar}, {Nakamura},
 {Narayan}, {Narayanan}, {Natarajan}, {Neri}, {Ni}, {Noutsos}, {Okino},
 {Olivares}, {Oyama}, {{\"O}zel}, {Palumbo}, {Patel}, {Pen}, {Pesce},
 {Pi{\'e}tu}, {Plambeck}, {PopStefanija}, {Porth}, {Prather},
 {Preciado-L{\'o}pez}, {Psaltis}, {Pu}, {Ramakrishnan}, {Rao}, {Rawlings},
 {Raymond}, {Rezzolla}, {Ripperda}, {Roelofs}, {Rogers}, {Ros}, {Rose},
 {Roshanineshat}, {Rottmann}, {Roy}, {Ruszczyk}, {Ryan}, {Rygl},
 {S{\'a}nchez}, {S{\'a}nchez-Arguelles}, {Sasada}, {Savolainen}, {Schloerb},
 {Schuster}, {Shao}, {Shen}, {Small}, {Sohn}, {SooHoo}, {Tazaki}, {Tiede},
 {Tilanus}, {Titus}, {Toma}, {Torne}, {Trent}, {Trippe}, {Tsuda}, {van
 Bemmel}, {van Langevelde}, {van Rossum}, {Wagner}, {Wardle}, {Weintroub},
 {Wex}, {Wharton}, {Wielgus}, {Wong}, {Wu}, {Young}, {Young}, {Younsi},
 {Yuan}, {Yuan}, {Zensus}, {Zhao}, {Zhao}, {Zhu}, {Farah}, {Meyer-Zhao},
 {Michalik}, {Nadolski}, {Nishioka}, {Pradel}, {Primiani}, {Souccar},
 {Vertatschitsch}, and {Yamaguchi}]{EHTM87PaperIV}
{Event Horizon Telescope Collaboration}.; {Akiyama}, K.; {Alberdi}, A.; {Alef},
 W.; {Asada}, K.; {Azulay}, R.; {Baczko}, A.K.; {Ball}, D.; {Balokovi{\'c}},
 M.; {Barrett}, J.; et al.
\newblock {First M87 Event Horizon Telescope Results. IV. Imaging the Central
 Supermassive Black Hole}.
\newblock {\em Astrophys. J. Lett.} {\bf 2019}, {\em 875}, L4. 
\newblock {{https://doi.org/10.3847/2041-8213/ab0e85}}.

\bibitem[{Event Horizon Telescope Collaboration}
 {et al.}(2019{\natexlab{e}}){Event Horizon Telescope Collaboration},
 {Akiyama}, {Alberdi}, {Alef}, {Asada}, {Azulay}, {Baczko}, {Ball},
 {Balokovi{\'c}}, {Barrett}, {Bintley}, {Blackburn}, {Boland}, {Bouman},
 {Bower}, {Bremer}, {Brinkerink}, {Brissenden}, {Britzen}, {Broderick},
 {Broguiere}, {Bronzwaer}, {Byun}, {Carlstrom}, {Chael}, {Chan}, {Chatterjee},
 {Chatterjee}, {Chen}, {Chen}, {Cho}, {Christian}, {Conway}, {Cordes}, {Crew},
 {Cui}, {Davelaar}, {De Laurentis}, {Deane}, {Dempsey}, {Desvignes}, {Dexter},
 {Doeleman}, {Eatough}, {Falcke}, {Fish}, {Fomalont}, {Fraga-Encinas},
 {Friberg}, {Fromm}, {G{\'o}mez}, {Galison}, {Gammie}, {Garc{\'\i}a},
 {Gentaz}, {Georgiev}, {Goddi}, {Gold}, {Gu}, {Gurwell}, {Hada}, {Hecht},
 {Hesper}, {Ho}, {Ho}, {Honma}, {Huang}, {Huang}, {Hughes}, {Ikeda}, {Inoue},
 {Issaoun}, {James}, {Jannuzi}, {Janssen}, {Jeter}, {Jiang}, {Johnson},
 {Jorstad}, {Jung}, {Karami}, {Karuppusamy}, {Kawashima}, {Keating},
 {Kettenis}, {Kim}, {Kim}, {Kim}, {Kino}, {Koay}, {Koch}, {Koyama}, {Kramer},
 {Kramer}, {Krichbaum}, {Kuo}, {Lauer}, {Lee}, {Li}, {Li}, {Lindqvist}, {Liu},
 {Liuzzo}, {Lo}, {Lobanov}, {Loinard}, {Lonsdale}, {Lu}, {MacDonald}, {Mao},
 {Markoff}, {Marrone}, {Marscher}, {Mart{\'\i}-Vidal}, {Matsushita},
 {Matthews}, {Medeiros}, {Menten}, {Mizuno}, {Mizuno}, {Moran}, {Moriyama},
 {Moscibrodzka}, {Mul{\ensuremath{\ddot{}}}ler}, {Nagai}, {Nagar}, {Nakamura},
 {Narayan}, {Narayanan}, {Natarajan}, {Neri}, {Ni}, {Noutsos}, {Okino},
 {Olivares}, {Oyama}, {{\"O}zel}, {Palumbo}, {Patel}, {Pen}, {Pesce},
 {Pi{\'e}tu}, {Plambeck}, {PopStefanija}, {Porth}, {Prather},
 {Preciado-L{\'o}pez}, {Psaltis}, {Pu}, {Ramakrishnan}, {Rao}, {Rawlings},
 {Raymond}, {Rezzolla}, {Ripperda}, {Roelofs}, {Rogers}, {Ros}, {Rose},
 {Roshanineshat}, {Rottmann}, {Roy}, {Ruszczyk}, {Ryan}, {Rygl},
 {S{\'a}nchez}, {S{\'a}nchez-Arguelles}, {Sasada}, {Savolainen}, {Schloerb},
 {Schuster}, {Shao}, {Shen}, {Small}, {Sohn}, {SooHoo}, {Tazaki}, {Tiede},
 {Tilanus}, {Titus}, {Toma}, {Torne}, {Trent}, {Trippe}, {Tsuda}, {van
 Bemmel}, {van Langevelde}, {van Rossum}, {Wagner}, {Wardle}, {Weintroub},
 {Wex}, {Wharton}, {Wielgus}, {Wong}, {Wu}, {Young}, {Young}, {Younsi},
 {Yuan}, {Yuan}, {Zensus}, {Zhao}, {Zhao}, {Zhu}, {Anczarski}, {Baganoff},
 {Eckart}, {Farah}, {Haggard}, {Meyer-Zhao}, {Michalik}, {Nadolski},
 {Neilsen}, {Nishioka}, {Nowak}, {Pradel}, {Primiani}, {Souccar},
 {Vertatschitsch}, {Yamaguchi}, and {Zhang}]{EHTM87PaperV}
{Event Horizon Telescope Collaboration}.; {Akiyama}, K.; {Alberdi}, A.; {Alef},
 W.; {Asada}, K.; {Azulay}, R.; {Baczko}, A.K.; {Ball}, D.; {Balokovi{\'c}},
 M.; {Barrett}, J.; et al.
\newblock {First M87 Event Horizon Telescope Results. V. Physical Origin of the
 Asymmetric Ring}.
\newblock {\em Astrophys. J. Lett.} {\bf 2019}, {\em 875}, L5. 
\newblock {{https://doi.org/10.3847/2041-8213/ab0f43}}.

\bibitem[{Event Horizon Telescope Collaboration}
 {et al.}(2019{\natexlab{f}}){Event Horizon Telescope Collaboration},
 {Akiyama}, {Alberdi}, {Alef}, {Asada}, {Azulay}, {Baczko}, {Ball},
 {Balokovi{\'c}}, {Barrett}, {Bintley}, {Blackburn}, {Boland}, {Bouman},
 {Bower}, {Bremer}, {Brinkerink}, {Brissenden}, {Britzen}, {Broderick},
 {Broguiere}, {Bronzwaer}, {Byun}, {Carlstrom}, {Chael}, {Chan}, {Chatterjee},
 {Chatterjee}, {Chen}, {Chen}, {Cho}, {Christian}, {Conway}, {Cordes}, {Crew},
 {Cui}, {Davelaar}, {De Laurentis}, {Deane}, {Dempsey}, {Desvignes}, {Dexter},
 {Doeleman}, {Eatough}, {Falcke}, {Fish}, {Fomalont}, {Fraga-Encinas},
 {Friberg}, {Fromm}, {G{\'o}mez}, {Galison}, {Gammie}, {Garc{\'\i}a},
 {Gentaz}, {Georgiev}, {Goddi}, {Gold}, {Gu}, {Gurwell}, {Hada}, {Hecht},
 {Hesper}, {Ho}, {Ho}, {Honma}, {Huang}, {Huang}, {Hughes}, {Ikeda}, {Inoue},
 {Issaoun}, {James}, {Jannuzi}, {Janssen}, {Jeter}, {Jiang}, {Johnson},
 {Jorstad}, {Jung}, {Karami}, {Karuppusamy}, {Kawashima}, {Keating},
 {Kettenis}, {Kim}, {Kim}, {Kim}, {Kino}, {Koay}, {Koch}, {Koyama}, {Kramer},
 {Kramer}, {Krichbaum}, {Kuo}, {Lauer}, {Lee}, {Li}, {Li}, {Lindqvist}, {Liu},
 {Liuzzo}, {Lo}, {Lobanov}, {Loinard}, {Lonsdale}, {Lu}, {MacDonald}, {Mao},
 {Markoff}, {Marrone}, {Marscher}, {Mart{\'\i}-Vidal}, {Matsushita},
 {Matthews}, {Medeiros}, {Menten}, {Mizuno}, {Mizuno}, {Moran}, {Moriyama},
 {Moscibrodzka}, {M{\"u}ller}, {Nagai}, {Nagar}, {Nakamura}, {Narayan},
 {Narayanan}, {Natarajan}, {Neri}, {Ni}, {Noutsos}, {Okino}, {Olivares},
 {Oyama}, {{\"O}zel}, {Palumbo}, {Patel}, {Pen}, {Pesce}, {Pi{\'e}tu},
 {Plambeck}, {PopStefanija}, {Porth}, {Prather}, {Preciado-L{\'o}pez},
 {Psaltis}, {Pu}, {Ramakrishnan}, {Rao}, {Rawlings}, {Raymond}, {Rezzolla},
 {Ripperda}, {Roelofs}, {Rogers}, {Ros}, {Rose}, {Roshanineshat}, {Rottmann},
 {Roy}, {Ruszczyk}, {Ryan}, {Rygl}, {S{\'a}nchez}, {S{\'a}nchez-Arguelles},
 {Sasada}, {Savolainen}, {Schloerb}, {Schuster}, {Shao}, {Shen}, {Small},
 {Sohn}, {SooHoo}, {Tazaki}, {Tiede}, {Tilanus}, {Titus}, {Toma}, {Torne},
 {Trent}, {Trippe}, {Tsuda}, {van Bemmel}, {van Langevelde}, {van Rossum},
 {Wagner}, {Wardle}, {Weintroub}, {Wex}, {Wharton}, {Wielgus}, {Wong}, {Wu},
 {Young}, {Young}, {Younsi}, {Yuan}, {Yuan}, {Zensus}, {Zhao}, {Zhao}, {Zhu},
 {Farah}, {Meyer-Zhao}, {Michalik}, {Nadolski}, {Nishioka}, {Pradel},
 {Primiani}, {Souccar}, {Vertatschitsch}, and {Yamaguchi}]{EHTM87PaperVI}
{Event Horizon Telescope Collaboration}.; {Akiyama}, K.; {Alberdi}, A.; {Alef},
 W.; {Asada}, K.; {Azulay}, R.; {Baczko}, A.K.; {Ball}, D.; {Balokovi{\'c}},
 M.; {Barrett}, J.; et al.
\newblock {First M87 Event Horizon Telescope Results. VI. The Shadow and Mass
 of the Central Black Hole}.
\newblock {\em Astrophys. J. Lett.} {\bf 2019}, {\em 875}, L6. 
\newblock {{https://doi.org/10.3847/2041-8213/ab1141}}.

\bibitem[{Event Horizon Telescope Collaboration}
 {et al.}(2021{\natexlab{a}}){Event Horizon Telescope Collaboration},
 {Akiyama}, {Algaba}, {Alberdi}, {Alef}, {Anantua}, {Asada}, {Azulay},
 {Baczko}, {Ball}, {Balokovi{\'c}}, {Barrett}, {Benson}, {Bintley},
 {Blackburn}, {Blundell}, {Boland}, {Bouman}, {Bower}, {Boyce}, {Bremer},
 {Brinkerink}, {Brissenden}, {Britzen}, {Broderick}, {Broguiere}, {Bronzwaer},
 {Byun}, {Carlstrom}, {Chael}, {Chan}, {Chatterjee}, {Chatterjee}, {Chen},
 {Chen}, {Chesler}, {Cho}, {Christian}, {Conway}, {Cordes}, {Crawford},
 {Crew}, {Cruz-Osorio}, {Cui}, {Davelaar}, {De Laurentis}, {Deane}, {Dempsey},
 {Desvignes}, {Dexter}, {Doeleman}, {Eatough}, {Falcke}, {Farah}, {Fish},
 {Fomalont}, {Ford}, {Fraga-Encinas}, {Freeman}, {Friberg}, {Fromm},
 {Fuentes}, {Galison}, {Gammie}, {Garc{\'\i}a}, {Gentaz}, {Georgiev}, {Goddi},
 {Gold}, {G{\'o}mez}, {G{\'o}mez-Ruiz}, {Gu}, {Gurwell}, {Hada}, {Haggard},
 {Hecht}, {Hesper}, {Ho}, {Ho}, {Honma}, {Huang}, {Huang}, {Hughes}, {Ikeda},
 {Inoue}, {Issaoun}, {James}, {Jannuzi}, {Janssen}, {Jeter}, {Jiang},
 {Jimenez-Rosales}, {Johnson}, {Jorstad}, {Jung}, {Karami}, {Karuppusamy},
 {Kawashima}, {Keating}, {Kettenis}, {Kim}, {Kim}, {Kim}, {Kim}, {Kino},
 {Koay}, {Kofuji}, {Koch}, {Koyama}, {Kramer}, {Kramer}, {Krichbaum}, {Kuo},
 {Lauer}, {Lee}, {Levis}, {Li}, {Li}, {Lindqvist}, {Lico}, {Lindahl}, {Liu},
 {Liu}, {Liuzzo}, {Lo}, {Lobanov}, {Loinard}, {Lonsdale}, {Lu}, {MacDonald},
 {Mao}, {Marchili}, {Markoff}, {Marrone}, {Marscher}, {Mart{\'\i}-Vidal},
 {Matsushita}, {Matthews}, {Medeiros}, {Menten}, {Mizuno}, {Mizuno}, {Moran},
 {Moriyama}, {Moscibrodzka}, {M{\"u}ller}, {Musoke}, {Mej{\'\i}as},
 {Michalik}, {Nadolski}, {Nagai}, {Nagar}, {Nakamura}, {Narayan}, {Narayanan},
 {Natarajan}, {Nathanail}, {Neilsen}, {Neri}, {Ni}, {Noutsos}, {Nowak},
 {Okino}, {Olivares}, {Ortiz-Le{\'o}n}, {Oyama}, {{\"O}zel}, {Palumbo},
 {Park}, {Patel}, {Pen}, {Pesce}, {Pi{\'e}tu}, {Plambeck}, {PopStefanija},
 {Porth}, {P{\"o}tzl}, {Prather}, {Preciado-L{\'o}pez}, {Psaltis}, {Pu},
 {Ramakrishnan}, {Rao}, {Rawlings}, {Raymond}, {Rezzolla}, {Ricarte},
 {Ripperda}, {Roelofs}, {Rogers}, {Ros}, {Rose}, {Roshanineshat}, {Rottmann},
 {Roy}, {Ruszczyk}, {Rygl}, {S{\'a}nchez}, {S{\'a}nchez-Arguelles}, {Sasada},
 {Savolainen}, {Schloerb}, {Schuster}, {Shao}, {Shen}, {Small}, {Sohn},
 {SooHoo}, {Sun}, {Tazaki}, {Tetarenko}, {Tiede}, {Tilanus}, {Titus}, {Toma},
 {Torne}, {Trent}, {Traianou}, {Trippe}, {van Bemmel}, {van Langevelde}, {van
 Rossum}, {Wagner}, {Ward-Thompson}, {Wardle}, {Weintroub}, {Wex}, {Wharton},
 {Wielgus}, {Wong}, {Wu}, {Yoon}, {Young}, {Young}, {Younsi}, {Yuan}, {Yuan},
 {Zensus}, {Zhao}, and {Zhao}]{EHTM87PaperVII}
{Event Horizon Telescope Collaboration}.; {Akiyama}, K.; {Algaba}, J.C.;
 {Alberdi}, A.; {Alef}, W.; {Anantua}, R.; {Asada}, K.; {Azulay}, R.;
 {Baczko}, A.K.; {Ball}, D.; et al.
\newblock {First M87 Event Horizon Telescope Results. VII. Polarization of the
 Ring}.
\newblock {\em Astrophys. J. Lett.} {\bf 2021}, {\em 910}, L12. 
\newblock {{https://doi.org/10.3847/2041-8213/abe71d}}.

\bibitem[{Event Horizon Telescope Collaboration}
 {et al.}(2021{\natexlab{b}}){Event Horizon Telescope Collaboration},
 {Akiyama}, {Algaba}, {Alberdi}, {Alef}, {Anantua}, {Asada}, {Azulay},
 {Baczko}, {Ball}, {Balokovi{\'c}}, {Barrett}, {Benson}, {Bintley},
 {Blackburn}, {Blundell}, {Boland}, {Bouman}, {Bower}, {Boyce}, {Bremer},
 {Brinkerink}, {Brissenden}, {Britzen}, {Broderick}, {Broguiere}, {Bronzwaer},
 {Byun}, {Carlstrom}, {Chael}, {Chan}, {Chatterjee}, {Chatterjee}, {Chen},
 {Chen}, {Chesler}, {Cho}, {Christian}, {Conway}, {Cordes}, {Crawford},
 {Crew}, {Cruz-Osorio}, {Cui}, {Davelaar}, {De Laurentis}, {Deane}, {Dempsey},
 {Desvignes}, {Dexter}, {Doeleman}, {Eatough}, {Falcke}, {Farah}, {Fish},
 {Fomalont}, {Ford}, {Fraga-Encinas}, {Friberg}, {Fromm}, {Fuentes},
 {Galison}, {Gammie}, {Garc{\'\i}a}, {Gelles}, {Gentaz}, {Georgiev}, {Goddi},
 {Gold}, {G{\'o}mez}, {G{\'o}mez-Ruiz}, {Gu}, {Gurwell}, {Hada}, {Haggard},
 {Hecht}, {Hesper}, {Himwich}, {Ho}, {Ho}, {Honma}, {Huang}, {Huang},
 {Hughes}, {Ikeda}, {Inoue}, {Issaoun}, {James}, {Jannuzi}, {Janssen},
 {Jeter}, {Jiang}, {Jimenez-Rosales}, {Johnson}, {Jorstad}, {Jung}, {Karami},
 {Karuppusamy}, {Kawashima}, {Keating}, {Kettenis}, {Kim}, {Kim}, {Kim},
 {Kim}, {Kino}, {Koay}, {Kofuji}, {Koch}, {Koyama}, {Kramer}, {Kramer},
 {Krichbaum}, {Kuo}, {Lauer}, {Lee}, {Levis}, {Li}, {Li}, {Lindqvist}, {Lico},
 {Lindahl}, {Liu}, {Liu}, {Liuzzo}, {Lo}, {Lobanov}, {Loinard}, {Lonsdale},
 {Lu}, {MacDonald}, {Mao}, {Marchili}, {Markoff}, {Marrone}, {Marscher},
 {Mart{\'\i}-Vidal}, {Matsushita}, {Matthews}, {Medeiros}, {Menten}, {Mizuno},
 {Mizuno}, {Moran}, {Moriyama}, {Moscibrodzka}, {M{\"u}ller}, {Musoke}, {Mus
 Mej{\'\i}as}, {Michalik}, {Nadolski}, {Nagai}, {Nagar}, {Nakamura},
 {Narayan}, {Narayanan}, {Natarajan}, {Nathanail}, {Neilsen}, {Neri}, {Ni},
 {Noutsos}, {Nowak}, {Okino}, {Olivares}, {Ortiz-Le{\'o}n}, {Oyama},
 {{\"O}zel}, {Palumbo}, {Park}, {Patel}, {Pen}, {Pesce}, {Pi{\'e}tu},
 {Plambeck}, {PopStefanija}, {Porth}, {P{\"o}tzl}, {Prather},
 {Preciado-L{\'o}pez}, {Psaltis}, {Pu}, {Ramakrishnan}, {Rao}, {Rawlings},
 {Raymond}, {Rezzolla}, {Ricarte}, {Ripperda}, {Roelofs}, {Rogers}, {Ros},
 {Rose}, {Roshanineshat}, {Rottmann}, {Roy}, {Ruszczyk}, {Rygl},
 {S{\'a}nchez}, {S{\'a}nchez-Arguelles}, {Sasada}, {Savolainen}, {Schloerb},
 {Schuster}, {Shao}, {Shen}, {Small}, {Sohn}, {SooHoo}, {Sun}, {Tazaki},
 {Tetarenko}, {Tiede}, {Tilanus}, {Titus}, {Toma}, {Torne}, {Trent},
 {Traianou}, {Trippe}, {van Bemmel}, {van Langevelde}, {van Rossum}, {Wagner},
 {Ward-Thompson}, {Wardle}, {Weintroub}, {Wex}, {Wharton}, {Wielgus}, {Wong},
 {Wu}, {Yoon}, {Young}, {Young}, {Younsi}, {Yuan}, {Yuan}, {Zensus}, {Zhao},
 and {Zhao}]{EHTM87PaperVIII}
{Event Horizon Telescope Collaboration}.; {Akiyama}, K.; {Algaba}, J.C.;
 {Alberdi}, A.; {Alef}, W.; {Anantua}, R.; {Asada}, K.; {Azulay}, R.;
 {Baczko}, A.K.; {Ball}, D.; et al.
\newblock {First M87 Event Horizon Telescope Results. VIII. Magnetic Field
 Structure near The Event Horizon}.
\newblock {\em Astrophys. J. Lett.} {\bf 2021}, {\em 910}, L13. 
\newblock {{https://doi.org/10.3847/2041-8213/abe4de}}.

\bibitem[{Event Horizon Telescope Collaboration}
 {et al.}(2022{\natexlab{a}}){Event Horizon Telescope Collaboration},
 {Akiyama}, {Alberdi}, {Alef}, {Algaba}, {Anantua}, {Asada}, {Azulay}, {Bach},
 {Baczko}, {Ball}, {Balokovi{\'c}}, {Barrett}, {Baub{\"o}ck}, {Benson},
 {Bintley}, {Blackburn}, {Blundell}, {Bouman}, {Bower}, {Boyce}, {Bremer},
 {Brinkerink}, {Brissenden}, {Britzen}, {Broderick}, {Broguiere}, {Bronzwaer},
 {Bustamante}, {Byun}, {Carlstrom}, {Ceccobello}, {Chael}, {Chan},
 {Chatterjee}, {Chatterjee}, {Chen}, {Chen}, {Cheng}, {Cho}, {Christian},
 {Conroy}, {Conway}, {Cordes}, {Crawford}, {Crew}, {Cruz-Osorio}, {Cui},
 {Davelaar}, {De Laurentis}, {Deane}, {Dempsey}, {Desvignes}, {Dexter},
 {Dhruv}, {Doeleman}, {Dougal}, {Dzib}, {Eatough}, {Emami}, {Falcke}, {Farah},
 {Fish}, {Fomalont}, {Ford}, {Fraga-Encinas}, {Freeman}, {Friberg}, {Fromm},
 {Fuentes}, {Galison}, {Gammie}, {Garc{\'\i}a}, {Gentaz}, {Georgiev}, {Goddi},
 {Gold}, {G{\'o}mez-Ruiz}, {G{\'o}mez}, {Gu}, {Gurwell}, {Hada}, {Haggard},
 {Haworth}, {Hecht}, {Hesper}, {Heumann}, {Ho}, {Ho}, {Honma}, {Huang},
 {Huang}, {Hughes}, {Ikeda}, {Impellizzeri}, {Inoue}, {Issaoun}, {James},
 {Jannuzi}, {Janssen}, {Jeter}, {Jiang}, {Jim{\'e}nez-Rosales}, {Johnson},
 {Jorstad}, {Joshi}, {Jung}, {Karami}, {Karuppusamy}, {Kawashima}, {Keating},
 {Kettenis}, {Kim}, {Kim}, {Kim}, {Kim}, {Kino}, {Koay}, {Kocherlakota},
 {Kofuji}, {Koch}, {Koyama}, {Kramer}, {Kramer}, {Krichbaum}, {Kuo}, {La
 Bella}, {Lauer}, {Lee}, {Lee}, {Leung}, {Levis}, {Li}, {Lico}, {Lindahl},
 {Lindqvist}, {Lisakov}, {Liu}, {Liu}, {Liuzzo}, {Lo}, {Lobanov}, {Loinard},
 {Lonsdale}, {Lu}, {Mao}, {Marchili}, {Markoff}, {Marrone}, {Marscher},
 {Mart{\'\i}-Vidal}, {Matsushita}, {Matthews}, {Medeiros}, {Menten},
 {Michalik}, {Mizuno}, {Mizuno}, {Moran}, {Moriyama}, {Moscibrodzka},
 {M{\"u}ller}, {Mus}, {Musoke}, {Myserlis}, {Nadolski}, {Nagai}, {Nagar},
 {Nakamura}, {Narayan}, {Narayanan}, {Natarajan}, {Nathanail}, {Navarro
 Fuentes}, {Neilsen}, {Neri}, {Ni}, {Noutsos}, {Nowak}, {Oh}, {Okino},
 {Olivares}, {Ortiz-Le{\'o}n}, {Oyama}, {{\"O}zel}, {Palumbo}, {Paraschos},
 {Park}, {Parsons}, {Patel}, {Pen}, {Pesce}, {Pi{\'e}tu}, {Plambeck},
 {PopStefanija}, {Porth}, {P{\"o}tzl}, {Prather}, {Preciado-L{\'o}pez},
 {Psaltis}, {Pu}, {Ramakrishnan}, {Rao}, {Rawlings}, {Raymond}, {Rezzolla},
 {Ricarte}, {Ripperda}, {Roelofs}, {Rogers}, {Ros}, {Romero-Ca{\ n}izales},
 {Roshanineshat}, {Rottmann}, {Roy}, {Ruiz}, {Ruszczyk}, {Rygl},
 {S{\'a}nchez}, {S{\'a}nchez-Arg{\"u}elles}, {S{\'a}nchez-Portal}, {Sasada},
 {Satapathy}, {Savolainen}, {Schloerb}, {Schonfeld}, {Schuster}, {Shao},
 {Shen}, {Small}, {Sohn}, {SooHoo}, {Souccar}, {Sun}, {Tazaki}, {Tetarenko},
 {Tiede}, {Tilanus}, {Titus}, {Torne}, {Traianou}, {Trent}, {Trippe}, {Turk},
 {van Bemmel}, {van Langevelde}, {van Rossum}, {Vos}, {Wagner},
 {Ward-Thompson}, {Wardle}, {Weintroub}, {Wex}, {Wharton}, {Wielgus}, {Wiik},
 {Witzel}, {Wondrak}, {Wong}, {Wu}, {Yamaguchi}, {Yoon}, {Young}, {Young},
 {Younsi}, {Yuan}, {Yuan}, {Zensus}, {Zhang}, {Zhao}, {Zhao}, {Agurto},
 {Allardi}, {Amestica}, {Araneda}, {Arriagada}, {Berghuis}, {Bertarini},
 {Berthold}, {Blanchard}, {Brown}, {C{\'a}rdenas}, {Cantzler}, {Caro},
 {Castillo-Dom{\'\i}nguez}, {Chan}, {Chang}, {Chang}, {Chang}, {Chang},
 {Chen}, {Chilson}, {Chuter}, {Ciechanowicz}, {Colin-Beltran}, {Coulson},
 {Crowley}, {Degenaar}, {Dornbusch}, {Dur{\'a}n}, {Everett}, {Faber},
 {Forster}, {Fuchs}, {Gale}, {Geertsema}, {Gonz{\'a}lez}, {Graham}, {Gueth},
 {Halverson}, {Han}, {Han}, {Hasegawa}, {Hern{\'a}ndez-Rebollar}, {Herrera},
 {Herrero-Illana}, {Heyminck}, {Hirota}, {Hoge}, {Hostler Schimpf}, {Howie},
 {Huang}, {Jiang}, {Jinchi}, {John}, {Kimura}, {Klein}, {Kubo}, {Kuroda},
 {Kwon}, {Lacasse}, {Laing}, {Leitch}, {Li}, {Liu}, {Liu}, {Lin}, {Lu},
 {Mac-Auliffe}, {Martin-Cocher}, {Matulonis}, {Maute}, {Messias},
 {Meyer-Zhao}, {Monta{\ n}a}, {Montenegro-Montes}, {Montgomerie}, {Moreno
 Nolasco}, {Muders}, {Nishioka}, {Norton}, {Nystrom}, {Ogawa}, {Olivares},
 {Oshiro}, {P{\'e}rez-Beaupuits}, {Parra}, {Phillips}, {Poirier}, {Pradel},
 {Qiu}, {Raffin}, {Rahlin}, {Ram{\'\i}rez}, {Ressler}, {Reynolds},
 {Rodr{\'\i}guez-Montoya}, {Saez-Madain}, {Santana}, {Shaw}, {Shirkey},
 {Silva}, {Snow}, {Sousa}, {Sridharan}, {Stahm}, {Stark}, {Test},
 {Torstensson}, {Venegas}, {Walther}, {Wei}, {White}, {Wieching}, {Wijnands},
 {Wouterloot}, {Yu}, {Yu}, {Zeballos}, and {Event Horizon Telescope
 Collaboration}]{EHTSgrAPaperI}
{Event Horizon Telescope Collaboration}.; {Akiyama}, K.; {Alberdi}, A.; {Alef},
 W.; {Algaba}, J.C.; {Anantua}, R.; {Asada}, K.; {Azulay}, R.; {Bach}, U.;
 {Baczko}, A.K.; et al.
\newblock {First Sagittarius A* Event Horizon Telescope Results. I. The Shadow
 of the Supermassive Black Hole in the Center of the Milky Way}.
\newblock {\em Astrophys. J. Lett.} {\bf 2022}, {\em 930}, L12.
\newblock {{https://doi.org/10.3847/2041-8213/ac6674}}.

\bibitem[{Event Horizon Telescope Collaboration}
 {et al.}(2022{\natexlab{b}}){Event Horizon Telescope Collaboration},
 {Akiyama}, {Alberdi}, {Alef}, {Algaba}, {Anantua}, {Asada}, {Azulay}, {Bach},
 {Baczko}, {Ball}, {Balokovi{\'c}}, {Barrett}, {Baub{\"o}ck}, {Benson},
 {Bintley}, {Blackburn}, {Blundell}, {Bouman}, {Bower}, {Boyce}, {Bremer},
 {Brinkerink}, {Brissenden}, {Britzen}, {Broderick}, {Broguiere}, {Bronzwaer},
 {Bustamante}, {Byun}, {Carlstrom}, {Ceccobello}, {Chael}, {Chan},
 {Chatterjee}, {Chatterjee}, {Chen}, {Chen}, {Cheng}, {Cho}, {Christian},
 {Conroy}, {Conway}, {Cordes}, {Crawford}, {Crew}, {Cruz-Osorio}, {Cui},
 {Davelaar}, {De Laurentis}, {Deane}, {Dempsey}, {Desvignes}, {Dexter},
 {Dhruv}, {Doeleman}, {Dougal}, {Dzib}, {Eatough}, {Emami}, {Falcke}, {Farah},
 {Fish}, {Fomalont}, {Ford}, {Fraga-Encinas}, {Freeman}, {Friberg}, {Fromm},
 {Fuentes}, {Galison}, {Gammie}, {Garc{\'\i}a}, {Gentaz}, {Georgiev}, {Goddi},
 {Gold}, {G{\'o}mez-Ruiz}, {G{\'o}mez}, {Gu}, {Gurwell}, {Hada}, {Haggard},
 {Haworth}, {Hecht}, {Hesper}, {Heumann}, {Ho}, {Ho}, {Honma}, {Huang},
 {Huang}, {Hughes}, {Ikeda}, {Impellizzeri}, {Inoue}, {Issaoun}, {James},
 {Jannuzi}, {Janssen}, {Jeter}, {Jiang}, {Jim{\'e}nez-Rosales}, {Johnson},
 {Jorstad}, {Joshi}, {Jung}, {Karami}, {Karuppusamy}, {Kawashima}, {Keating},
 {Kettenis}, {Kim}, {Kim}, {Kim}, {Kim}, {Kino}, {Koay}, {Kocherlakota},
 {Kofuji}, {Koch}, {Koyama}, {Kramer}, {Kramer}, {Krichbaum}, {Kuo}, {La
 Bella}, {Lauer}, {Lee}, {Lee}, {Leung}, {Levis}, {Li}, {Lico}, {Lindahl},
 {Lindqvist}, {Lisakov}, {Liu}, {Liu}, {Liuzzo}, {Lo}, {Lobanov}, {Loinard},
 {Lonsdale}, {Lu}, {Mao}, {Marchili}, {Markoff}, {Marrone}, {Marscher},
 {Mart{\'\i}-Vidal}, {Matsushita}, {Matthews}, {Medeiros}, {Menten},
 {Michalik}, {Mizuno}, {Mizuno}, {Moran}, {Moriyama}, {Moscibrodzka},
 {M{\"u}ller}, {Mus}, {Musoke}, {Myserlis}, {Nadolski}, {Nagai}, {Nagar},
 {Nakamura}, {Narayan}, {Narayanan}, {Natarajan}, {Nathanail}, {Navarro
 Fuentes}, {Neilsen}, {Neri}, {Ni}, {Noutsos}, {Nowak}, {Oh}, {Okino},
 {Olivares}, {Ortiz-Le{\'o}n}, {Oyama}, {{\"O}zel}, {Palumbo}, {Paraschos},
 {Park}, {Parsons}, {Patel}, {Pen}, {Pesce}, {Pi{\'e}tu}, {Plambeck},
 {PopStefanija}, {Porth}, {P{\"o}tzl}, {Prather}, {Preciado-L{\'o}pez},
 {Psaltis}, {Pu}, {Ramakrishnan}, {Rao}, {Rawlings}, {Raymond}, {Rezzolla},
 {Ricarte}, {Ripperda}, {Roelofs}, {Rogers}, {Ros}, {Romero-Ca{\ n}izales},
 {Roshanineshat}, {Rottmann}, {Roy}, {Ruiz}, {Ruszczyk}, {Rygl},
 {S{\'a}nchez}, {S{\'a}nchez-Arg{\"u}elles}, {S{\'a}nchez-Portal}, {Sasada},
 {Satapathy}, {Savolainen}, {Schloerb}, {Schonfeld}, {Schuster}, {Shao},
 {Shen}, {Small}, {Sohn}, {SooHoo}, {Souccar}, {Sun}, {Tazaki}, {Tetarenko},
 {Tiede}, {Tilanus}, {Titus}, {Torne}, {Traianou}, {Trent}, {Trippe}, {Turk},
 {van Bemmel}, {van Langevelde}, {van Rossum}, {Vos}, {Wagner},
 {Ward-Thompson}, {Wardle}, {Weintroub}, {Wex}, {Wharton}, {Wielgus}, {Wiik},
 {Witzel}, {Wondrak}, {Wong}, {Wu}, {Yamaguchi}, {Yoon}, {Young}, {Young},
 {Younsi}, {Yuan}, {Yuan}, {Zensus}, {Zhang}, {Zhao}, {Zhao}, {Agurto},
 {Araneda}, {Arriagada}, {Bertarini}, {Berthold}, {Blanchard}, {Brown},
 {C{\'a}rdenas}, {Cantzler}, {Caro}, {Chuter}, {Ciechanowicz}, {Coulson},
 {Crowley}, {Degenaar}, {Dornbusch}, {Dur{\'a}n}, {Forster}, {Geertsema},
 {Gonz{\'a}lez}, {Graham}, {Gueth}, {Han}, {Herrera}, {Herrero-Illana},
 {Heyminck}, {Hoge}, {Huang}, {Jiang}, {John}, {Klein}, {Kubo}, {Kuroda},
 {Kwon}, {Laing}, {Liu}, {Liu}, {Mac-Auliffe}, {Martin-Cocher}, {Matulonis},
 {Messias}, {Meyer-Zhao}, {Montenegro-Montes}, {Montgomerie}, {Muders},
 {Nishioka}, {Norton}, {Olivares}, {P{\'e}rez-Beaupuits}, {Parra}, {Poirier},
 {Pradel}, {Raffin}, {Ram{\'\i}rez}, {Reynolds}, {Saez-Madain}, {Santana},
 {Silva}, {Sousa}, {Stahm}, {Torstensson}, {Venegas}, {Walther}, {Wieching},
 {Wijnands}, {Wouterloot}, and {Event Horizon Telescope
 Collaboration}]{EHTSgrAPaperII}
{Event Horizon Telescope Collaboration}.; {Akiyama}, K.; {Alberdi}, A.; {Alef},
 W.; {Algaba}, J.C.; {Anantua}, R.; {Asada}, K.; {Azulay}, R.; {Bach}, U.;
 {Baczko}, A.K.; et al.
\newblock {First Sagittarius A* Event Horizon Telescope Results. II. EHT and
 Multiwavelength Observations, Data Processing, and Calibration}.
\newblock {\em Astrophys. J. Lett.} {\bf 2022}, {\em 930}, L13.
\newblock {{https://doi.org/10.3847/2041-8213/ac6675}}.

\bibitem[{Event Horizon Telescope Collaboration}
 {et al.}(2022{\natexlab{c}}){Event Horizon Telescope Collaboration},
 {Akiyama}, {Alberdi}, {Alef}, {Algaba}, {Anantua}, {Asada}, {Azulay}, {Bach},
 {Baczko}, {Ball}, {Balokovi{\'c}}, {Barrett}, {Baub{\"o}ck}, {Benson},
 {Bintley}, {Blackburn}, {Blundell}, {Bouman}, {Bower}, {Boyce}, {Bremer},
 {Brinkerink}, {Brissenden}, {Britzen}, {Broderick}, {Broguiere}, {Bronzwaer},
 {Bustamante}, {Byun}, {Carlstrom}, {Ceccobello}, {Chael}, {Chan},
 {Chatterjee}, {Chatterjee}, {Chen}, {Chen}, {Cheng}, {Cho}, {Christian},
 {Conroy}, {Conway}, {Cordes}, {Crawford}, {Crew}, {Cruz-Osorio}, {Cui},
 {Davelaar}, {De Laurentis}, {Deane}, {Dempsey}, {Desvignes}, {Dexter},
 {Dhruv}, {Doeleman}, {Dougal}, {Dzib}, {Eatough}, {Emami}, {Falcke}, {Farah},
 {Fish}, {Fomalont}, {Ford}, {Fraga-Encinas}, {Freeman}, {Friberg}, {Fromm},
 {Fuentes}, {Galison}, {Gammie}, {Garc{\'\i}a}, {Gentaz}, {Georgiev}, {Goddi},
 {Gold}, {G{\'o}mez-Ruiz}, {G{\'o}mez}, {Gu}, {Gurwell}, {Hada}, {Haggard},
 {Haworth}, {Hecht}, {Hesper}, {Heumann}, {Ho}, {Ho}, {Honma}, {Huang},
 {Huang}, {Hughes}, {Ikeda}, {Impellizzeri}, {Inoue}, {Issaoun}, {James},
 {Jannuzi}, {Janssen}, {Jeter}, {Jiang}, {Jim{\'e}nez-Rosales}, {Johnson},
 {Jorstad}, {Joshi}, {Jung}, {Karami}, {Karuppusamy}, {Kawashima}, {Keating},
 {Kettenis}, {Kim}, {Kim}, {Kim}, {Kim}, {Kino}, {Koay}, {Kocherlakota},
 {Kofuji}, {Koch}, {Koyama}, {Kramer}, {Kramer}, {Krichbaum}, {Kuo}, {La
 Bella}, {Lauer}, {Lee}, {Lee}, {Leung}, {Levis}, {Li}, {Lico}, {Lindahl},
 {Lindqvist}, {Lisakov}, {Liu}, {Liu}, {Liuzzo}, {Lo}, {Lobanov}, {Loinard},
 {Lonsdale}, {Lu}, {Mao}, {Marchili}, {Markoff}, {Marrone}, {Marscher},
 {Mart{\'\i}-Vidal}, {Matsushita}, {Matthews}, {Medeiros}, {Menten},
 {Michalik}, {Mizuno}, {Mizuno}, {Moran}, {Moriyama}, {Moscibrodzka},
 {M{\"u}ller}, {Mus}, {Musoke}, {Myserlis}, {Nadolski}, {Nagai}, {Nagar},
 {Nakamura}, {Narayan}, {Narayanan}, {Natarajan}, {Nathanail}, {Navarro
 Fuentes}, {Neilsen}, {Neri}, {Ni}, {Noutsos}, {Nowak}, {Oh}, {Okino},
 {Olivares}, {Ortiz-Le{\'o}n}, {Oyama}, {{\"O}zel}, {Palumbo}, {Paraschos},
 {Park}, {Parsons}, {Patel}, {Pen}, {Pesce}, {Pi{\'e}tu}, {Plambeck},
 {PopStefanija}, {Porth}, {P{\"o}tzl}, {Prather}, {Preciado-L{\'o}pez},
 {Psaltis}, {Pu}, {Ramakrishnan}, {Rao}, {Rawlings}, {Raymond}, {Rezzolla},
 {Ricarte}, {Ripperda}, {Roelofs}, {Rogers}, {Ros}, {Romero-Ca{\ n}izales},
 {Roshanineshat}, {Rottmann}, {Roy}, {Ruiz}, {Ruszczyk}, {Rygl},
 {S{\'a}nchez}, {S{\'a}nchez-Arg{\"u}elles}, {S{\'a}nchez-Portal}, {Sasada},
 {Satapathy}, {Savolainen}, {Schloerb}, {Schonfeld}, {Schuster}, {Shao},
 {Shen}, {Small}, {Sohn}, {SooHoo}, {Souccar}, {Sun}, {Tazaki}, {Tetarenko},
 {Tiede}, {Tilanus}, {Titus}, {Torne}, {Traianou}, {Trent}, {Trippe}, {Turk},
 {van Bemmel}, {van Langevelde}, {van Rossum}, {Vos}, {Wagner},
 {Ward-Thompson}, {Wardle}, {Weintroub}, {Wex}, {Wharton}, {Wielgus}, {Wiik},
 {Witzel}, {Wondrak}, {Wong}, {Wu}, {Yamaguchi}, {Yoon}, {Young}, {Young},
 {Younsi}, {Yuan}, {Yuan}, {Zensus}, {Zhang}, {Zhao}, {Zhao}, and {Event
 Horizon Telescope Collaboration}]{EHTSgrAPaperIII}
{Event Horizon Telescope Collaboration}.; {Akiyama}, K.; {Alberdi}, A.; {Alef},
 W.; {Algaba}, J.C.; {Anantua}, R.; {Asada}, K.; {Azulay}, R.; {Bach}, U.;
 {Baczko}, A.K.; et al.
\newblock {First Sagittarius A* Event Horizon Telescope Results. III. Imaging
 of the Galactic Center Supermassive Black Hole}.
\newblock {\em Astrophys. J. Lett.} {\bf 2022}, {\em 930}, L14.
\newblock {{https://doi.org/10.3847/2041-8213/ac6429}}.

\bibitem[{Event Horizon Telescope Collaboration}
 {et al.}(2022{\natexlab{d}}){Event Horizon Telescope Collaboration},
 {Akiyama}, {Alberdi}, {Alef}, {Algaba}, {Anantua}, {Asada}, {Azulay}, {Bach},
 {Baczko}, {Ball}, {Balokovi{\'c}}, {Barrett}, {Baub{\"o}ck}, {Benson},
 {Bintley}, {Blackburn}, {Blundell}, {Bouman}, {Bower}, {Boyce}, {Bremer},
 {Brinkerink}, {Brissenden}, {Britzen}, {Broderick}, {Broguiere}, {Bronzwaer},
 {Bustamante}, {Byun}, {Carlstrom}, {Ceccobello}, {Chael}, {Chan},
 {Chatterjee}, {Chatterjee}, {Chen}, {Chen}, {Cheng}, {Cho}, {Christian},
 {Conroy}, {Conway}, {Cordes}, {Crawford}, {Crew}, {Cruz-Osorio}, {Cui},
 {Davelaar}, {De Laurentis}, {Deane}, {Dempsey}, {Desvignes}, {Dexter},
 {Dhruv}, {Doeleman}, {Dougal}, {Dzib}, {Eatough}, {Emami}, {Falcke}, {Farah},
 {Fish}, {Fomalont}, {Ford}, {Fraga-Encinas}, {Freeman}, {Friberg}, {Fromm},
 {Fuentes}, {Galison}, {Gammie}, {Garc{\'\i}a}, {Gentaz}, {Georgiev}, {Goddi},
 {Gold}, {G{\'o}mez-Ruiz}, {G{\'o}mez}, {Gu}, {Gurwell}, {Hada}, {Haggard},
 {Haworth}, {Hecht}, {Hesper}, {Heumann}, {Ho}, {Ho}, {Honma}, {Huang},
 {Huang}, {Hughes}, {Ikeda}, {Impellizzeri}, {Inoue}, {Issaoun}, {James},
 {Jannuzi}, {Janssen}, {Jeter}, {Jiang}, {Jim{\'e}nez-Rosales}, {Johnson},
 {Jorstad}, {Joshi}, {Jung}, {Karami}, {Karuppusamy}, {Kawashima}, {Keating},
 {Kettenis}, {Kim}, {Kim}, {Kim}, {Kim}, {Kino}, {Koay}, {Kocherlakota},
 {Kofuji}, {Koch}, {Koyama}, {Kramer}, {Kramer}, {Krichbaum}, {Kuo}, {La
 Bella}, {Lauer}, {Lee}, {Lee}, {Leung}, {Levis}, {Li}, {Lico}, {Lindahl},
 {Lindqvist}, {Lisakov}, {Liu}, {Liu}, {Liuzzo}, {Lo}, {Lobanov}, {Loinard},
 {Lonsdale}, {Lu}, {Mao}, {Marchili}, {Markoff}, {Marrone}, {Marscher},
 {Mart{\'\i}-Vidal}, {Matsushita}, {Matthews}, {Medeiros}, {Menten},
 {Michalik}, {Mizuno}, {Mizuno}, {Moran}, {Moriyama}, {Moscibrodzka},
 {M{\"u}ller}, {Mus}, {Musoke}, {Myserlis}, {Nadolski}, {Nagai}, {Nagar},
 {Nakamura}, {Narayan}, {Narayanan}, {Natarajan}, {Nathanail}, {Navarro
 Fuentes}, {Neilsen}, {Neri}, {Ni}, {Noutsos}, {Nowak}, {Oh}, {Okino},
 {Olivares}, {Ortiz-Le{\'o}n}, {Oyama}, {Palumbo}, {Paraschos}, {Park},
 {Parsons}, {Patel}, {Pen}, {Pesce}, {Pi{\'e}tu}, {Plambeck}, {PopStefanija},
 {Porth}, {P{\"o}tzl}, {Prather}, {Preciado-L{\'o}pez}, {Pu}, {Ramakrishnan},
 {Rao}, {Rawlings}, {Raymond}, {Rezzolla}, {Ricarte}, {Ripperda}, {Roelofs},
 {Rogers}, {Ros}, {Romero-Ca{\ n}izales}, {Roshanineshat}, {Rottmann}, {Roy},
 {Ruiz}, {Ruszczyk}, {Rygl}, {S{\'a}nchez}, {S{\'a}nchez-Arg{\"u}elles},
 {S{\'a}nchez-Portal}, {Sasada}, {Satapathy}, {Savolainen}, {Schloerb},
 {Schonfeld}, {Schuster}, {Shao}, {Shen}, {Small}, {Sohn}, {SooHoo},
 {Souccar}, {Sun}, {Tazaki}, {Tetarenko}, {Tiede}, {Tilanus}, {Titus},
 {Torne}, {Traianou}, {Trent}, {Trippe}, {Turk}, {van Bemmel}, {van
 Langevelde}, {van Rossum}, {Vos}, {Wagner}, {Ward-Thompson}, {Wardle},
 {Weintroub}, {Wex}, {Wharton}, {Wielgus}, {Wiik}, {Witzel}, {Wondrak},
 {Wong}, {Wu}, {Yamaguchi}, {Yoon}, {Young}, {Young}, {Younsi}, {Yuan},
 {Yuan}, {Zensus}, {Zhang}, {Zhao}, {Zhao}, {Chang}, and {Event Horizon
 Telescope Collaboration}]{EHTSgrAPaperIV}
{Event Horizon Telescope Collaboration}.; {Akiyama}, K.; {Alberdi}, A.; {Alef},
 W.; {Algaba}, J.C.; {Anantua}, R.; {Asada}, K.; {Azulay}, R.; {Bach}, U.;
 {Baczko}, A.K.; et al.
\newblock {First Sagittarius A* Event Horizon Telescope Results. IV.
 Variability, Morphology, and Black Hole Mass}.
\newblock {\em Astrophys. J. Lett.} {\bf 2022}, {\em 930}, L15.
\newblock {{https://doi.org/10.3847/2041-8213/ac6736}}.

\bibitem[{Event Horizon Telescope Collaboration}
 {et al.}(2022{\natexlab{e}}){Event Horizon Telescope Collaboration},
 {Akiyama}, {Alberdi}, {Alef}, {Algaba}, {Anantua}, {Asada}, {Azulay}, {Bach},
 {Baczko}, {Ball}, {Balokovi{\'c}}, {Barrett}, {Baub{\"o}ck}, {Benson},
 {Bintley}, {Blackburn}, {Blundell}, {Bouman}, {Bower}, {Boyce}, {Bremer},
 {Brinkerink}, {Brissenden}, {Britzen}, {Broderick}, {Broguiere}, {Bronzwaer},
 {Bustamante}, {Byun}, {Carlstrom}, {Ceccobello}, {Chael}, {Chan},
 {Chatterjee}, {Chatterjee}, {Chen}, {Chen}, {Cheng}, {Cho}, {Christian},
 {Conroy}, {Conway}, {Cordes}, {Crawford}, {Crew}, {Cruz-Osorio}, {Cui},
 {Davelaar}, {De Laurentis}, {Deane}, {Dempsey}, {Desvignes}, {Dexter},
 {Dhruv}, {Doeleman}, {Dougal}, {Dzib}, {Eatough}, {Emami}, {Falcke}, {Farah},
 {Fish}, {Fomalont}, {Ford}, {Fraga-Encinas}, {Freeman}, {Friberg}, {Fromm},
 {Fuentes}, {Galison}, {Gammie}, {Garc{\'\i}a}, {Gentaz}, {Georgiev}, {Goddi},
 {Gold}, {G{\'o}mez-Ruiz}, {G{\'o}mez}, {Gu}, {Gurwell}, {Hada}, {Haggard},
 {Haworth}, {Hecht}, {Hesper}, {Heumann}, {Ho}, {Ho}, {Honma}, {Huang},
 {Huang}, {Hughes}, {Ikeda}, {Impellizzeri}, {Inoue}, {Issaoun}, {James},
 {Jannuzi}, {Janssen}, {Jeter}, {Jiang}, {Jim{\'e}nez-Rosales}, {Johnson},
 {Jorstad}, {Joshi}, {Jung}, {Karami}, {Karuppusamy}, {Kawashima}, {Keating},
 {Kettenis}, {Kim}, {Kim}, {Kim}, {Kim}, {Kino}, {Koay}, {Kocherlakota},
 {Kofuji}, {Koch}, {Koyama}, {Kramer}, {Kramer}, {Krichbaum}, {Kuo}, {La
 Bella}, {Lauer}, {Lee}, {Lee}, {Leung}, {Levis}, {Li}, {Lico}, {Lindahl},
 {Lindqvist}, {Lisakov}, {Liu}, {Liu}, {Liuzzo}, {Lo}, {Lobanov}, {Loinard},
 {Lonsdale}, {Lu}, {Mao}, {Marchili}, {Markoff}, {Marrone}, {Marscher},
 {Mart{\'\i}-Vidal}, {Matsushita}, {Matthews}, {Medeiros}, {Menten},
 {Michalik}, {Mizuno}, {Mizuno}, {Moran}, {Moriyama}, {Moscibrodzka},
 {M{\"u}ller}, {Mus}, {Musoke}, {Myserlis}, {Nadolski}, {Nagai}, {Nagar},
 {Nakamura}, {Narayan}, {Narayanan}, {Natarajan}, {Nathanail}, {Navarro
 Fuentes}, {Neilsen}, {Neri}, {Ni}, {Noutsos}, {Nowak}, {Oh}, {Okino},
 {Olivares}, {Ortiz-Le{\'o}n}, {Oyama}, {{\"O}zel}, {Palumbo}, {Paraschos},
 {Park}, {Parsons}, {Patel}, {Pen}, {Pesce}, {Pi{\'e}tu}, {Plambeck},
 {PopStefanija}, {Porth}, {P{\"o}tzl}, {Prather}, {Preciado-L{\'o}pez},
 {Psaltis}, {Pu}, {Ramakrishnan}, {Rao}, {Rawlings}, {Raymond}, {Rezzolla},
 {Ricarte}, {Ripperda}, {Roelofs}, {Rogers}, {Ros}, {Romero-Ca{\ n}izales},
 {Roshanineshat}, {Rottmann}, {Roy}, {Ruiz}, {Ruszczyk}, {Rygl},
 {S{\'a}nchez}, {S{\'a}nchez-Arg{\"u}elles}, {S{\'a}nchez-Portal}, {Sasada},
 {Satapathy}, {Savolainen}, {Schloerb}, {Schonfeld}, {Schuster}, {Shao},
 {Shen}, {Small}, {Sohn}, {SooHoo}, {Souccar}, {Sun}, {Tazaki}, {Tetarenko},
 {Tiede}, {Tilanus}, {Titus}, {Torne}, {Traianou}, {Trent}, {Trippe}, {Turk},
 {van Bemmel}, {van Langevelde}, {van Rossum}, {Vos}, {Wagner},
 {Ward-Thompson}, {Wardle}, {Weintroub}, {Wex}, {Wharton}, {Wielgus}, {Wiik},
 {Witzel}, {Wondrak}, {Wong}, {Wu}, {Yamaguchi}, {Yoon}, {Young}, {Young},
 {Younsi}, {Yuan}, {Yuan}, {Zensus}, {Zhang}, {Zhao}, {Zhao}, {Chan}, {Qiu},
 {Ressler}, {White}, and {Event Horizon Telescope
 Collaboration}]{EHTSgrAPaperV}
{Event Horizon Telescope Collaboration}.; {Akiyama}, K.; {Alberdi}, A.; {Alef},
 W.; {Algaba}, J.C.; {Anantua}, R.; {Asada}, K.; {Azulay}, R.; {Bach}, U.;
 {Baczko}, A.K.; et al.
\newblock {First Sagittarius A* Event Horizon Telescope Results. V. Testing
 Astrophysical Models of the Galactic Center Black Hole}.
\newblock {\em Astrophys. J. Lett.} {\bf 2022}, {\em 930}, L16.
\newblock {{https://doi.org/10.3847/2041-8213/ac6672}}.

\bibitem[{Event Horizon Telescope Collaboration}
 {et al.}(2022{\natexlab{f}}){Event Horizon Telescope Collaboration},
 {Akiyama}, {Alberdi}, {Alef}, {Algaba}, {Anantua}, {Asada}, {Azulay}, {Bach},
 {Baczko}, {Ball}, {Balokovi{\'c}}, {Barrett}, {Baub{\"o}ck}, {Benson},
 {Bintley}, {Blackburn}, {Blundell}, {Bouman}, {Bower}, {Boyce}, {Bremer},
 {Brinkerink}, {Brissenden}, {Britzen}, {Broderick}, {Broguiere}, {Bronzwaer},
 {Bustamante}, {Byun}, {Carlstrom}, {Ceccobello}, {Chael}, {Chan},
 {Chatterjee}, {Chatterjee}, {Chen}, {Chen}, {Cheng}, {Cho}, {Christian},
 {Conroy}, {Conway}, {Cordes}, {Crawford}, {Crew}, {Cruz-Osorio}, {Cui},
 {Davelaar}, {De Laurentis}, {Deane}, {Dempsey}, {Desvignes}, {Dexter},
 {Dhruv}, {Doeleman}, {Dougal}, {Dzib}, {Eatough}, {Emami}, {Falcke}, {Farah},
 {Fish}, {Fomalont}, {Ford}, {Fraga-Encinas}, {Freeman}, {Friberg}, {Fromm},
 {Fuentes}, {Galison}, {Gammie}, {Garc{\'\i}a}, {Gentaz}, {Georgiev}, {Goddi},
 {Gold}, {G{\'o}mez-Ruiz}, {G{\'o}mez}, {Gu}, {Gurwell}, {Hada}, {Haggard},
 {Haworth}, {Hecht}, {Hesper}, {Heumann}, {Ho}, {Ho}, {Honma}, {Huang},
 {Huang}, {Hughes}, {Ikeda}, {Impellizzeri}, {Inoue}, {Issaoun}, {James},
 {Jannuzi}, {Janssen}, {Jeter}, {Jiang}, {Jim{\'e}nez-Rosales}, {Johnson},
 {Jorstad}, {Joshi}, {Jung}, {Karami}, {Karuppusamy}, {Kawashima}, {Keating},
 {Kettenis}, {Kim}, {Kim}, {Kim}, {Kim}, {Kino}, {Koay}, {Kocherlakota},
 {Kofuji}, {Koch}, {Koyama}, {Kramer}, {Kramer}, {Krichbaum}, {Kuo}, {La
 Bella}, {Lauer}, {Lee}, {Lee}, {Leung}, {Levis}, {Li}, {Lico}, {Lindahl},
 {Lindqvist}, {Lisakov}, {Liu}, {Liu}, {Liuzzo}, {Lo}, {Lobanov}, {Loinard},
 {Lonsdale}, {Lu}, {Mao}, {Marchili}, {Markoff}, {Marrone}, {Marscher},
 {Mart{\'\i}-Vidal}, {Matsushita}, {Matthews}, {Medeiros}, {Menten},
 {Michalik}, {Mizuno}, {Mizuno}, {Moran}, {Moriyama}, {Moscibrodzka},
 {M{\"u}ller}, {Mus}, {Musoke}, {Myserlis}, {Nadolski}, {Nagai}, {Nagar},
 {Nakamura}, {Narayan}, {Narayanan}, {Natarajan}, {Nathanail}, {Navarro
 Fuentes}, {Neilsen}, {Neri}, {Ni}, {Noutsos}, {Nowak}, {Oh}, {Okino},
 {Olivares}, {Ortiz-Le{\'o}n}, {Oyama}, {{\"O}zel}, {Palumbo}, {Paraschos},
 {Park}, {Parsons}, {Patel}, {Pen}, {Pesce}, {Pi{\'e}tu}, {Plambeck},
 {PopStefanija}, {Porth}, {P{\"o}tzl}, {Prather}, {Preciado-L{\'o}pez},
 {Psaltis}, {Pu}, {Ramakrishnan}, {Rao}, {Rawlings}, {Raymond}, {Rezzolla},
 {Ricarte}, {Ripperda}, {Roelofs}, {Rogers}, {Ros}, {Romero-Ca{\ n}izales},
 {Roshanineshat}, {Rottmann}, {Roy}, {Ruiz}, {Ruszczyk}, {Rygl},
 {S{\'a}nchez}, {S{\'a}nchez-Arg{\"u}elles}, {S{\'a}nchez-Portal}, {Sasada},
 {Satapathy}, {Savolainen}, {Schloerb}, {Schonfeld}, {Schuster}, {Shao},
 {Shen}, {Small}, {Sohn}, {SooHoo}, {Souccar}, {Sun}, {Tazaki}, {Tetarenko},
 {Tiede}, {Tilanus}, {Titus}, {Torne}, {Traianou}, {Trent}, {Trippe}, {Turk},
 {van Bemmel}, {van Langevelde}, {van Rossum}, {Vos}, {Wagner},
 {Ward-Thompson}, {Wardle}, {Weintroub}, {Wex}, {Wharton}, {Wielgus}, {Wiik},
 {Witzel}, {Wondrak}, {Wong}, {Wu}, {Yamaguchi}, {Yoon}, {Young}, {Young},
 {Younsi}, {Yuan}, {Yuan}, {Zensus}, {Zhang}, {Zhao}, {Zhao}, and {Event
 Horizon Telescope Collaboration}]{EHTSgrAPaperVI}
{Event Horizon Telescope Collaboration}.; {Akiyama}, K.; {Alberdi}, A.; {Alef},
 W.; {Algaba}, J.C.; {Anantua}, R.; {Asada}, K.; {Azulay}, R.; {Bach}, U.;
 {Baczko}, A.K.; et al.
\newblock {First Sagittarius A* Event Horizon Telescope Results. VI. Testing
 the Black Hole Metric}.
\newblock {\em Astrophys. J. Lett.} {\bf 2022}, {\em 930}, L17.
\newblock {{https://doi.org/10.3847/2041-8213/ac6756}}.

\bibitem[{Doeleman} {et al.}(2019{\natexlab{a}}){Doeleman}, {Akiyama},
 {Blackburn}, {Bouman}, {Bower}, {Broderick}, {Chael}, {Fish}, {Johnson},
 {Lonsdale}, {Krichbaum}, {Palumbo}, {Pesce}, {Raymond}, {Weintroub}, and
 {Wielgus}]{Doeleman2019a}
{Doeleman}, S.; {Akiyama}, K.; {Blackburn}, L.; {Bouman}, K.L.; {Bower}, G.C.;
 {Broderick}, A.E.; {Chael}, A.; {Fish}, V.L.; {Johnson}, M.D.; {Lonsdale},
 C.J.; et al.
\newblock {Black Hole Physics on Horizon Scales}.
\newblock {\em Bull. Am. Astron. Soc.} {\bf 2019}, {\em 51}, 537.

\bibitem[{Doeleman} {et al.}(2019{\natexlab{b}}){Doeleman}, {Blackburn},
 {Dexter}, {Gomez}, {Johnson}, {Palumbo}, {Weintroub}, {Farah}, {Fish},
 {Loinard}, {Lonsdale}, {Narayanan}, {Patel}, {Pesce}, {Raymond}, {Tilanus},
 {Wielgus}, {Akiyama}, {Bower}, {Broderick}, {Deane}, {Fromm}, {Gammie},
 {Gold}, {Janssen}, {Kawashima}, {Krichbaum}, {Marrone}, {Matthews}, {Mizuno},
 {Rezzolla}, {Roelofs}, {Ros}, {Savolainen}, {Yuan}, {Zhao}, {Blackburn},
 {Doeleman}, {Dexter}, {Gomez}, {Johnson}, {Palumbo}, {Weintroub}, {Farah},
 {Fish}, {Loinard}, {Lonsdale}, {Narayanan}, {Patel}, {Pesce}, {Raymond},
 {Tilanus}, {Wielgus}, {Akiyama}, {Bower}, {Broderick}, {Deane}, {Fromm},
 {Gammie}, {Gold}, {Janssen}, {Kawashima}, {Krichbaum}, {Marrone}, {Matthews},
 {Mizuno}, {Rezzolla}, {Roelofs}, {Ros}, {Savolainen}, {Yuan}, and
 {Zhao}]{Doeleman2019b}
{Doeleman}, S.; {Blackburn}, L.; {Dexter}, J.; {Gomez}, J.L.; {Johnson}, M.D.;
 {Palumbo}, D.C.; {Weintroub}, J.; {Farah}, J.R.; {Fish}, V.; {Loinard}, L.;
 et al.
\newblock {Studying Black Holes on Horizon Scales with VLBI Ground Arrays}. {\emph{arXiv} \textbf{2019}, arXiv:1909.01411.} 

\bibitem[{Raymond} {et al.}(2021){Raymond}, {Palumbo}, {Paine}, {Blackburn},
 {C{\'o}rdova Rosado}, {Doeleman}, {Farah}, {Johnson}, {Roelofs}, {Tilanus},
 and {Weintroub}]{Raymond2021}
{Raymond}, A.W.; {Palumbo}, D.; {Paine}, S.N.; {Blackburn}, L.; {C{\'o}rdova
 Rosado}, R.; {Doeleman}, S.S.; {Farah}, J.R.; {Johnson}, M.D.; {Roelofs}, F.;
 {Tilanus}, R.P.J.; et al.
\newblock {Evaluation of New Submillimeter VLBI Sites for the Event Horizon
 Telescope}.
\newblock {\em Astrophys. J. Suppl. Ser.} {\bf 2021}, {\em 253}, 5. 
\newblock {{https://doi.org/10.3847/1538-3881/abc3c3}}.

\bibitem[{Selina} {et al.}(2018){Selina}, {Murphy}, {McKinnon}, {Beasley},
 {Butler}, {Carilli}, {Clark}, {Durand}, {Erickson}, {Grammer}, {Hiriart},
 {Jackson}, {Kent}, {Mason}, {Morgan}, {Ojeda}, {Rosero}, {Shillue},
 {Sturgis}, and {Urbain}]{Selina2018}
{Selina}, R.J.; {Murphy}, E.J.; {McKinnon}, M.; {Beasley}, A.; {Butler}, B.;
 {Carilli}, C.; {Clark}, B.; {Durand}, S.; {Erickson}, A.; {Grammer}, W.;
 et al.
\newblock {The ngVLA Reference Design}. In \emph{Science with a Next Generation Very Large Array}; Astronomical Society of the Pacific Conference Series; {Murphy}, E., Ed.; {ASP Monograph 7; ASP: San Francisco, CA, USA, }2018; Volume 517, p. 15. 



\bibitem[{Murphy} {et al.}(2018){Murphy}, {Bolatto}, {Chatterjee}, {Casey},
 {Chomiuk}, {Dale}, {de Pater}, {Dickinson}, {Francesco}, {Hallinan},
 {Isella}, {Kohno}, {Kulkarni}, {Lang}, {Lazio}, {Leroy}, {Loinard},
 {Maccarone}, {Matthews}, {Osten}, {Reid}, {Riechers}, {Sakai}, {Walter}, and
 {Wilner}]{Murphy2018}
{Murphy}, E.J.; {Bolatto}, A.; {Chatterjee}, S.; {Casey}, C.M.; {Chomiuk}, L.;
 {Dale}, D.; {de Pater}, I.; {Dickinson}, M.; {Francesco}, J.D.; {Hallinan},
 G.; et al.
\newblock {The ngVLA Science Case and Associated Science Requirements}. In \emph{Science with a Next Generation Very Large Array}; Astronomical Society of the Pacific Conference Series; {Murphy}, E., Ed.; {ASP Monograph 7; ASP: San Francisco, CA, USA, }2018; Volume 517, p. 3. 

\bibitem[{Issaoun} {et al.}(2022){Issaoun}, {Pesce}, {Roelofs}, and
 et al.]{Issaoun2022}
{Issaoun}, S.; {Pesce}, D.W.; {Roelofs}, F.; {Chael}, A.; {Dodson}, R.; {Rioja}, M.J.; {Akiyama}, K.; {Aran}, R.; {Blackburn}, L.; {Doeleman}, S.S.; et al.
\newblock {Enabling transformational ngEHT science via the inclusion of 86 GHz
 capabilities}.
\newblock {\em Galaxies} {\bf 2022}, \emph{submitted}. %

\bibitem[{Pesce} {et al.}(2021){Pesce}, {Palumbo}, {Narayan}, {Blackburn},
 {Doeleman}, {Johnson}, {Ma}, {Nagar}, {Natarajan}, and {Ricarte}]{Pesce2021}
{Pesce}, D.W.; {Palumbo}, D.C.M.; {Narayan}, R.; {Blackburn}, L.; {Doeleman},
 S.S.; {Johnson}, M.D.; {Ma}, C.P.; {Nagar}, N.M.; {Natarajan}, P.; {Ricarte},
 A.
\newblock {Toward Determining the Number of Observable Supermassive Black Hole
 Shadows}.
\newblock {\em Astrophys. J.} {\bf 2021}, {\em 923}, 260. 
\newblock {{https://doi.org/10.3847/1538-4357/ac2eb5}}.

\bibitem[{Kauffmann} {et al.}(2022){Kauffmann}, {Rajagopalan}, and
 et al.]{Kauffmann2022}
{Kauffmann}, J.; {Rajagopalan}, G.; {Akiyama}, K.; {Fish}, V.L.; {Lonsdale}, C.J.; {Matthews}, L.D.; Pillai, T.
\newblock {The Haystack Telescope as an Astronomical Instrument}.
\newblock {\em Galaxies} {\bf 2022}, \emph{submitted}.

\bibitem[{Kawabe} {et al.}(2016){Kawabe}, {Kohno}, {Tamura}, {Takekoshi},
 {Oshima}, and {Ishii}]{Kawabe2016}
{Kawabe}, R.; {Kohno}, K.; {Tamura}, Y.; {Takekoshi}, T.; {Oshima}, T.;
 {Ishii}, S.
\newblock {New 50-m-class single-dish telescope: Large Submillimeter Telescope
 (LST)}.
\newblock In {\em Ground-Based and Airborne Telescopes VI}; {Hall}, H.J., {Gilmozzi}, R., {Marshall}, H.K., Eds.; {Society of Photo-Optical Instrumentation Engineers (SPIE) Conference Series}; SPIE: Bellingham, WA, USA, 2016; Volume 9906, pp. 779--790.
\newblock {{https://doi.org/10.1117/12.2232202}}.

\bibitem[{Kohno} {et al.}(2020){Kohno}, {Kawabe}, {Tamura}, {Endo},
 {Baselmans}, {Karatsu}, {Inoue}, {Moriwaki}, {Hayatsu}, {Yoshida},
 {Yoshimura}, {Hatsukade}, {Umehata}, {Oshima}, {Takekoshi}, {Taniguchi},
 {Klaassen}, {Mroczkowski}, {Cicone}, {Bertoldi}, {Dannerbauer}, and
 {Tosaki}]{Kohno2020}
{Kohno}, K.; {Kawabe}, R.; {Tamura}, Y.; {Endo}, A.; {Baselmans}, J.J.A.;
 {Karatsu}, K.; {Inoue}, A.K.; {Moriwaki}, K.; {Hayatsu}, N.H.; {Yoshida}, N.;
 et al.
\newblock {Large format imaging spectrograph for the Large Submillimeter
 Telescope (LST)}.
\newblock In \emph{Millimeter, Submillimeter, and Far-Infrared Detectors and Instrumentation for Astronomy X}; Society of Photo-Optical Instrumentation
 Engineers (SPIE) Conference Series; SPIE: Bellingham, WA, USA, 2020; Volume 11453, pp. 128--138.
\newblock {{https://doi.org/10.1117/12.2561238}}.

\bibitem[{Klaassen} {et al.}(2020){Klaassen}, {Mroczkowski}, {Cicone},
 {Hatziminaoglou}, {Sartori}, {De Breuck}, {Bryan}, {Dicker}, {Duran},
 {Groppi}, {Kaercher}, {Kawabe}, {Kohno}, and {Geach}]{Klaassen2020}
{Klaassen}, P.D.; {Mroczkowski}, T.K.; {Cicone}, C.; {Hatziminaoglou}, E.;
 {Sartori}, S.; {De Breuck}, C.; {Bryan}, S.; {Dicker}, S.R.; {Duran}, C.;
 {Groppi}, C.; et al.
\newblock {The Atacama Large Aperture Submillimeter Telescope (AtLAST)}.
\newblock In \emph{Ground-Based and Airborne Telescopes VIII}; Society of Photo-Optical Instrumentation
 Engineers (SPIE) Conference Series; SPIE: Bellingham, WA, USA, 2020, Volume 11445, pp. 544--563. 
\newblock {{https://doi.org/10.1117/12.2561315}}.

\bibitem[{Ramasawmy} {et al.}(2022){Ramasawmy}, {Klaassen}, {Cicone},
 {Mroczkowski}, {Chen}, {Cornish}, {da Cunha}, {Hatziminaoglou}, {Johnstone},
 {Liu}, {Perrott}, {Schimek}, {Stanke}, and {Wedemeyer}]{Ramasawmy2022}
{Ramasawmy}, J.; {Klaassen}, P.D.; {Cicone}, C.; {Mroczkowski}, T.K.; {Chen},
 C.C.; {Cornish}, T.; {da Cunha}, E.; {Hatziminaoglou}, E.; {Johnstone}, D.;
 {Liu}, D.; et al.
\newblock {The Atacama Large Aperture Submillimetre Telescope: Key science
 drivers}.
\newblock In \emph{Millimeter, Submillimeter, and Far-Infrared Detectors and Instrumentation for Astronomy XI}; {Zmuidzinas}, J., {Gao},
 J.R., Eds.; {Society of Photo-Optical Instrumentation
 Engineers (SPIE) Conference Series}; SPIE: Bellingham, WA, USA, 2022, Volume 12190, pp. 112--130.
\newblock {{https://doi.org/10.1117/12.2627505}}.

\bibitem[{Ukita} and {Tsuboi}(1994)]{Ukita1994}
{Ukita}, N.; {Tsuboi}, M.
\newblock {A 45-meter telescope with a surface accuracy of 65 {${\upmu}$}m.}
\newblock {\em IEEE Proc.} {\bf 1994}, {\em 82}, 725--733.

\bibitem[{Ezawa} {et al.}(2008){Ezawa}, {Kohno}, {Kawabe}, {Yamamoto},
 {Inoue}, {Iwashita}, {Matsuo}, {Okuda}, {Oshima}, {Sakai}, {Tanaka},
 {Yamaguchi}, {Wilson}, {Yun}, {Aretxaga}, {Hughes}, {Austermann}, {Perera},
 {Scott}, {Bronfman}, and {Cortes}]{Ezawa2008}
{Ezawa}, H.; {Kohno}, K.; {Kawabe}, R.; {Yamamoto}, S.; {Inoue}, H.;
 {Iwashita}, H.; {Matsuo}, H.; {Okuda}, T.; {Oshima}, T.; {Sakai}, T.; et al.
\newblock {New achievements of ASTE: the Atacama Submillimeter Telescope
 Experiment}.
\newblock In \emph{Ground-Based and Airborne Telescopes II}; {Stepp}, L.M., {Gilmozzi}, R., Eds.; {Society of
 Photo-Optical Instrumentation Engineers (SPIE) Conference Series}; SPIE: Bellingham, WA, USA, 2008, Volume 7012, pp. 88--96.
\newblock {{https://doi.org/10.1117/12.789652}}.

\bibitem[{Tamura} {et al.}(2020){Tamura}, {Kawabe}, {Fukasaku}, {Kimura},
 {Ueda}, {Taniguchi}, {Okada}, {Ogawa}, {Hashimoto}, {Minamidani},
 {Kawaguchi}, {Kuno}, {Togami}, {Hagimoto}, {Nakano}, {Matsuda}, {Okumura},
 {Nakamura}, {Kurita}, {Takekoshi}, {Oshima}, {Onishi}, and
 {Kohno}]{Tamura2020}
{Tamura}, Y.; {Kawabe}, R.; {Fukasaku}, Y.; {Kimura}, K.; {Ueda}, T.;
 {Taniguchi}, A.; {Okada}, N.; {Ogawa}, H.; {Hashimoto}, I.; {Minamidani}, T.;
 et al.
\newblock {Wavefront sensor for millimeter/submillimeter-wave adaptive optics
 based on aperture-plane interferometry}.
\newblock In \emph{Ground-Based and Airborne Telescopes VIII}; Society of Photo-Optical Instrumentation
 Engineers (SPIE) Conference Series; SPIE: Bellingham, WA, USA, 2020; Volume 11445, pp. 350--358.
\newblock {{https://doi.org/10.1117/12.2561885}}.

\bibitem[{Issaoun} {et al.}(2019){Issaoun}, {Johnson}, {Blackburn},
 {Brinkerink}, {Mo{\'s}cibrodzka}, {Chael}, {Goddi}, {Mart{\'\i}-Vidal},
 {Wagner}, {Doeleman}, {Falcke}, {Krichbaum}, {Akiyama}, {Bach}, {Bouman},
 {Bower}, {Broderick}, {Cho}, {Crew}, {Dexter}, {Fish}, {Gold}, {G{\'o}mez},
 {Hada}, {Hern{\'a}ndez-G{\'o}mez}, {Jan{\ss}en}, {Kino}, {Kramer}, {Loinard},
 {Lu}, {Markoff}, {Marrone}, {Matthews}, {Moran}, {M{\"u}ller}, {Roelofs},
 {Ros}, {Rottmann}, {Sanchez}, {Tilanus}, {de Vicente}, {Wielgus}, {Zensus},
 and {Zhao}]{Issaoun2019}
{Issaoun}, S.; {Johnson}, M.D.; {Blackburn}, L.; {Brinkerink}, C.D.;
 {Mo{\'s}cibrodzka}, M.; {Chael}, A.; {Goddi}, C.; {Mart{\'\i}-Vidal}, I.;
 {Wagner}, J.; {Doeleman}, S.S.; et al.
\newblock {The Size, Shape, and Scattering of Sagittarius A* at 86 GHz: First
 VLBI with ALMA}.
\newblock {\em Astrophys. J.} {\bf 2019}, {\em 871}, 30.
\newblock {{https://doi.org/10.3847/1538-4357/aaf732}}.

\bibitem[{Issaoun} {et al.}(2021){Issaoun}, {Johnson}, {Blackburn},
 {Broderick}, {Tiede}, {Wielgus}, {Doeleman}, {Falcke}, {Akiyama}, {Bower},
 {Brinkerink}, {Chael}, {Cho}, {G{\'o}mez}, {Hern{\'a}ndez-G{\'o}mez},
 {Hughes}, {Kino}, {Krichbaum}, {Liuzzo}, {Loinard}, {Markoff}, {Marrone},
 {Mizuno}, {Moran}, {Pidopryhora}, {Ros}, {Rygl}, {Shen}, and
 {Wagner}]{Issaoun2021}
{Issaoun}, S.; {Johnson}, M.D.; {Blackburn}, L.; {Broderick}, A.; {Tiede}, P.;
 {Wielgus}, M.; {Doeleman}, S.S.; {Falcke}, H.; {Akiyama}, K.; {Bower}, G.C.;
 et al.
\newblock {Persistent Non-Gaussian Structure in the Image of Sagittarius A* at
 86 GHz}.
\newblock {\em Astrophys. J.} {\bf 2021}, {\em 915}, 99.
\newblock {{https://doi.org/10.3847/1538-4357/ac00b0}}.

\bibitem[{Okino} {et al.}(2021){Okino}, {Akiyama}, {Asada}, {G{\'o}mez},
 {Hada}, {Honma}, {Krichbaum}, {Kino}, {Nagai}, {Nakamura}, {Bach},
 {Blackburn}, {Bouman}, {Chael}, {Crew}, {Doeleman}, {Fish}, {Gabuzda},
 {Goddi}, {Issaoun}, {Johnson}, {Jorstad}, {Koyama}, {Lonsdale},
 {Mart{\'\i}-Vidal}, {Matthews}, {Mizuno}, {Moriyama}, {Pu}, {Ros},
 {Savolainen}, {Tazaki}, {Wagner}, {Wielgus}, and {Zensus}]{Okino2021}
{Okino}, H.; {Akiyama}, K.; {Asada}, K.; {G{\'o}mez}, J.L.; {Hada}, K.;
 {Honma}, M.; {Krichbaum}, T.P.; {Kino}, M.; {Nagai}, H.; {Nakamura}, M.;
 et al.
\newblock {Collimation of the relativistic jet in the quasar 3C 273}.
\newblock {\em arXiv} {\bf 2021}, arXiv:2112.12233.

\bibitem[{Zhao} {et al.}(2022){Zhao}, {G{\'o}mez}, {Fuentes}, {Krichbaum},
 {Traianou}, {Lico}, {Cho}, {Ros}, {Komossa}, {Akiyama}, {Asada}, {Blackburn},
 {Britzen}, {Bruni}, {Crew}, {Dahale}, {Dey}, {Gold}, {Gopakumar}, {Issaoun},
 {Janssen}, {Jorstad}, {Kim}, {Koay}, {Kovalev}, {Koyama}, {Lobanov},
 {Loinard}, {Lu}, {Markoff}, {Marscher}, {Mart{\'\i}-Vidal}, {Mizuno}, {Park},
 {Savolainen}, and {Toscano}]{Zhao2022}
{Zhao}, G.Y.; {G{\'o}mez}, J.L.; {Fuentes}, A.; {Krichbaum}, T.P.; {Traianou},
 E.; {Lico}, R.; {Cho}, I.; {Ros}, E.; {Komossa}, S.; {Akiyama}, K.; et al.
\newblock {Unraveling the Innermost Jet Structure of OJ 287 with the First GMVA
 + ALMA Observations}.
\newblock {\em Astrophys. J.} {\bf 2022}, {\em 932}, 72.
\newblock {{https://doi.org/10.3847/1538-4357/ac6b9c}}.

\bibitem[{Fish} {et al.}(2011){Fish}, {Doeleman}, {Beaudoin}, {Blundell},
 {Bolin}, {Bower}, {Chamberlin}, {Freund}, {Friberg}, {Gurwell}, {Honma},
 {Inoue}, {Krichbaum}, {Lamb}, {Marrone}, {Moran}, {Oyama}, {Plambeck},
 {Primiani}, {Rogers}, {Smythe}, {SooHoo}, {Strittmatter}, {Tilanus}, {Titus},
 {Weintroub}, {Wright}, {Woody}, {Young}, and {Ziurys}]{Fish2011}
{Fish}, V.L.; {Doeleman}, S.S.; {Beaudoin}, C.; {Blundell}, R.; {Bolin}, D.E.;
 {Bower}, G.C.; {Chamberlin}, R.; {Freund}, R.; {Friberg}, P.; {Gurwell},
 M.A.; et al.
\newblock {1.3 mm Wavelength VLBI of Sagittarius A*: Detection of Time-variable
 Emission on Event Horizon Scales}.
\newblock {\em Astrophys. J. Lett.} {\bf 2011}, {\em 727}, L36.
\newblock {{https://doi.org/10.1088/2041-8205/727/2/L36}}.

\bibitem[{Johnson} {et al.}(2015){Johnson}, {Fish}, {Doeleman}, {Marrone},
 {Plambeck}, {Wardle}, {Akiyama}, {Asada}, {Beaudoin}, {Blackburn},
 {Blundell}, {Bower}, {Brinkerink}, {Broderick}, {Cappallo}, {Chael}, {Crew},
 {Dexter}, {Dexter}, {Freund}, {Friberg}, {Gold}, {Gurwell}, {Ho}, {Honma},
 {Inoue}, {Kosowsky}, {Krichbaum}, {Lamb}, {Loeb}, {Lu}, {MacMahon},
 {McKinney}, {Moran}, {Narayan}, {Primiani}, {Psaltis}, {Rogers}, {Rosenfeld},
 {SooHoo}, {Tilanus}, {Titus}, {Vertatschitsch}, {Weintroub}, {Wright},
 {Young}, {Zensus}, and {Ziurys}]{Johnson2015}
{Johnson}, M.D.; {Fish}, V.L.; {Doeleman}, S.S.; {Marrone}, D.P.; {Plambeck},
 R.L.; {Wardle}, J.F.C.; {Akiyama}, K.; {Asada}, K.; {Beaudoin}, C.;
 {Blackburn}, L.; et al.
\newblock {Resolved magnetic-field structure and variability near the event
 horizon of Sagittarius A*}.
\newblock {\em Science} {\bf 2015}, {\em 350}, 1242--1245.
\newblock {{https://doi.org/10.1126/science.aac7087}}.

\bibitem[{Asaki} {et al.}(1996){Asaki}, {Saito}, {Kawabe}, {Morita}, and
 {Sasao}]{Asaki1996}
{Asaki}, Y.; {Saito}, M.; {Kawabe}, R.; {Morita}, K.; {Sasao}, T.
\newblock {A Phase Compensation Experiment with the Paired Antenna Method}. {\emph{Radio Sci.} \textbf{1996}, \emph{31}, 464.} 

\bibitem[{Dodson} and {Rioja}(2009)]{Dodson2009}
{Dodson}, R.; {Rioja}, M.J.
\newblock {VLBA Scientific Memorandum n. 31: Astrometric calibration of mm-VLBI
 using ``Source/Frequency Phase Referenced'' observations}.
\newblock {\em arXiv} {\bf 2009}, arXiv:0910.1159.

\bibitem[{Rioja} and {Dodson}(2011)]{Rioja2011}
{Rioja}, M.; {Dodson}, R.
\newblock {High-precision Astrometric Millimeter Very Long Baseline
 Interferometry Using a New Method for Atmospheric Calibration}.
\newblock {\em Astrophys. J.} {\bf 2011}, {\em 141}, 114.
\newblock {{https://doi.org/10.1088/0004-6256/141/4/114}}.

\bibitem[{Rioja} {et al.}(2022){Rioja}, {Dodson}, {Asaki}, and
 et al.]{Rioja2022}
{Rioja}, M.J.; {Dodson}, R.; {Asaki}, Y.
\newblock {The Transformational Power of Frequency Phase Transfer Methods for ngEHT}.
\newblock {\em Galaxies} {\bf 2022}, \emph{submitted}.

\bibitem[{Carlson} {et al.}(2020){Carlson}, {Pleasance}, {Gunaratne}, and
 {Vrcic}]{Carlson2020}
{Carlson}, B.; {Pleasance}, M.; {Gunaratne}, T.; {Vrcic}, S.
\newblock \emph{Study Report: NRC TALON Frequency Slice Architecture Correlator/Beamformer (AT.CBF) for ALMA}; {ALMA Memo 617}; {ALMA: New York, NY, USA,} {2020}.

\bibitem[{Matthews} {et al.}(2018){Matthews}, {Crew}, {Doeleman}, {Lacasse},
 {Saez}, {Alef}, {Akiyama}, {Amestica}, {Anderson}, {Barkats}, {Baudry},
 {Brogui{\`e}re}, {Escoffier}, {Fish}, {Greenberg}, {Hecht}, {Hiriart},
 {Hirota}, {Honma}, {Ho}, {Impellizzeri}, {Inoue}, {Kohno}, {Lopez},
 {Mart{\'\i}-Vidal}, {Messias}, {Meyer-Zhao}, {Mora-Klein}, {Nagar},
 {Nishioka}, {Oyama}, {Pankratius}, {Perez}, {Phillips}, {Pradel}, {Rottmann},
 {Roy}, {Ruszczyk}, {Shillue}, {Suzuki}, and {Treacy}]{Matthews2018}
{Matthews}, L.D.; {Crew}, G.B.; {Doeleman}, S.S.; {Lacasse}, R.; {Saez}, A.F.;
 {Alef}, W.; {Akiyama}, K.; {Amestica}, R.; {Anderson}, J.M.; {Barkats}, D.A.;
 et al.
\newblock {The ALMA Phasing System: A Beamforming Capability for
 Ultra-high-resolution Science at (Sub)Millimeter Wavelengths}.
\newblock {\em Publ. Astron. Soc. Pac.} {\bf 2018}, {\em 130}, 15002.
\newblock {{https://doi.org/10.1088/1538-3873/aa9c3d}}.

\bibitem[{Crew} {et al.}(2022){Crew}, {Goddi}, {Matthews}, and
 et al.]{Crew2022}
{Crew}, G.B.; {Goddi}, C.; {Matthews}, L.D.; {Rottmann}, H.; {Saez}, A.; {Mart\'\i-Vidal}, I.
\newblock {A Characterization of the ALMA Phasing System at 345 GHz}.
\newblock {\em Publ. Astron. Soc. Pac.} {\bf 2022}, \emph{submitted}.

\end{thebibliography}
\end{document}